\documentclass[aps,prb,twocolumn,showpacs,superscriptaddress]{revtex4-1}
\usepackage{amsmath}
\usepackage{graphicx}
\usepackage{subfigure}
\usepackage{color}
\begin{document}
\title{Dynamic magnetic susceptibility and electrical detection
of ferromagnetic resonance}
\author{Yin Zhang}
\author{X. S. Wang}
\author{H. Y. Yuan}
\affiliation{Physics Department, The Hong Kong University of Science and
Technology, Clear Water Bay, Kowloon, Hong Kong}
\affiliation{HKUST Shenzhen Research Institute, Shenzhen 518057, China}
\author{S. S. Kang}
\affiliation{School of Physics, National Key Laboratory of Crystal Materials,
Shandong University, Jinan, Shandong 250100, China}
\author{H. W. Zhang}
\affiliation{School of Microelectronics and Solid-State Electronics,
University of Electronic Science and Technology of China, Chengdu,
Sichuan 610054, China}
\author{X. R. Wang}
\email{[Corresponding author:]phxwan@ust.hk}
\affiliation{Physics Department, The Hong Kong University of Science and
Technology, Clear Water Bay, Kowloon, Hong Kong}
\affiliation{HKUST Shenzhen Research Institute, Shenzhen 518057, China}
\date{\today}

\begin{abstract}
The dynamic magnetic susceptibility of magnetic materials near
ferromagnetic resonance (FMR) is very important in interpreting
dc-voltage obtained in an electrical detection of FMR. Based on the
causality principle and the assumption that the usual microwave
absorption lineshape around FMR is Lorentzian, general forms of dynamic
magnetic susceptibility of an arbitrary sample and the corresponding
dc-voltage lineshapes of the electrical detection of FMR are obtained.
Our main findings are: 1) The dynamic magnetic susceptibility is not
a Polder tensor for a material with an arbitrary magnetic anisotropy.
Two off-diagonal matrix elements of the tensor near FMR are not in
general opposite to each other. However, the linear response coefficient
of the magnetization to the total radio frequency (rf) field (sum of the
applied external rf field and the internal rf field due to the precessing
magnetization, a quantity cannot be measured directly) is a Polder tensor.
This may explain why two off-diagonal susceptibility matrix elements are
always assumed to be opposite to each other in analyses.
2) The frequency dependence of dynamic magnetic susceptibility near FMR
is fully characterized by six {\it real} numbers while its field dependence
is fully characterized by seven {\it real} numbers. 3) A recipe of how to
determine these numbers by standard microwave absorption measurements
for a sample with an arbitrary magnetic anisotropy is proposed.
Our results allow one to unambiguously separate the contribution of the
anisotropic magnetoresistance to dc-voltage signals from that of the
anomalous Hall effect. With these results, one can reliably extract the
information of spin pumping and the inverse spin Hall effect, and
determine the spin-Hall angle. 4) The field-dependence of the
susceptibility matrix elements at a fixed microwave frequency may
have several peaks when the effective magnetic field is not a
monotonic function of the applied magnetic field. In contrast,
the frequency-dependence of the susceptibility matrix elements
at a fixed applied magnetic field has only one FMR peak.
Furthermore, in the case that resonance frequency is not sensitive to the
externally applied static magnetic field, the field dependence of
matrix elements of the dynamic magnetic susceptibility,
as well as dc-voltage, may have another non-resonance broad peak.
Thus, one should be careful in interpreting observed peaks.
\end{abstract}
\maketitle

\section{Introduction}
Ferromagnetic resonance (FMR) is an important phenomenon which can be
used to probe magnetic properties of ferromagnetic materials besides
many other applications \cite{kittel,suhl,tannenwald,seidel,lax,farle}.
Microwave absorption is usually measured in traditional FMR
experiments, and the absorption lineshape around FMR is a Lorentzian
function of microwave frequency for a fixed static magnetic field.
The peak position and peak width can be used to determine magnetic
anisotropy and the Gilbert damping coefficient \cite{mizukami,mzwu}.
In recent years, the electrical detection of FMR, in which a dc-voltage
is measured on a sample around FMR, becomes very popular due to the
high accuracy of dc-voltage measurements and only microns sample size
needed \cite{juretschke,egan,saitoh,costache,hoffmann,buhrman,azevedo,
hfding,ohno,ysgui,harder,tulapurkar,sankey,goennenwein,yamaguchi}.
This technique has been used by many groups \cite{saitoh,costache,
hoffmann,buhrman,azevedo,hfding,ohno} in recent years to extract spin
pumping and the spin Hall angle that measures the strength of both
the spin Hall effect (SHE) and the inverse spin Hall effect (ISHE).
However, the experimentally extracted material parameters show a large
discrepancy for the same materials with similar experimental setup.
For example, the spin Hall angle of Pt obtained by different groups differs
from each other by two orders of magnitude \cite{hoffmann,buhrman,azevedo}
due to different interpretations of the dc-voltage signal.

The dc-voltage in an electrical detection of FMR can come from two sources.
One is the generalized Ohm's law in which the well-known anisotropic
magnetoresistance (AMR) and anomalous Hall effect (AHE) couple the
magnetization motion with the electric current (see the discussion
below) \cite{juretschke, egan, harder}. This coupling between the
magnetization and microwave-induced electric current results in the
so-called spin rectification effect \cite{harder} that gives rise to a
detectable dc-voltage inside a magnetic layer or multilayer at FMR.
The other dc-voltage source is the ISHE: a spin current converts into
a transverse charge current via ISHE inside a multilayer, and the
charge current, in turn, gives rise to a dc-voltage.
In the popular spin-Hall-angle experiments \cite{saitoh,costache,hoffmann,
hfding,ohno}, the spin current comes from spin pumping by magnetization
precession at FMR \cite{bauer, kajiwara}.
To use the electrical detection of FMR as a probe of material properties,
one needs to separate different dc-voltage sources.
A common way is the symmetry analysis of dc-voltage lineshapes.
The ISHE contribution to dc-voltage is normally assumed to have a
symmetric lineshape \cite{saitoh,costache,hoffmann,buhrman,azevedo,
hfding,ohno}, resembling microwave absorption curves.
The AMR contribution to dc-voltage can be symmetric \cite{ohno},
antisymmetric \cite{hoffmann} or asymmetric \cite{azevedo, hfding}
although  resistance normally links to energy dissipation.
On the other hand, the AHE contribution can be antisymmetric
\cite{saitoh,ohno} or vanishing small \cite{hoffmann,azevedo,hfding}
because the Hall effect does not necessarily involve energy
dissipation, depending on material properties and experimental setup.
So far, most analyses \cite{saitoh,costache,hoffmann,azevedo,
hfding,harder} assume that the static response of magnetization is
always along the external magnetic field so that the microwave-induced
magnetization precession is around the external static magnetic field.
Furthermore, the two off-diagonal matrix elements of the
dynamic magnetic susceptibility are assumed to be opposite to each other.
However, both assumptions are questionable for layer/multilayer samples.
Interfacial spin-orbit interactions and/or other interactions may
modify the magnetic anisotropy of a film \cite{sakurai,tsujikawa,zli}.
As we shall see below, the dynamic magnetic susceptibility of a sample
with non-zero magnetic anisotropy is not a Polder tensor in general.
For a given experimental setup, the dc-voltage due to AMR and AHE is
very sensitive to the dynamic magnetic susceptibility since the spin
rectification effect comes from the phase lag of radio frequency (rf)
magnetization precession and rf current. Thus, the dynamic magnetic
susceptibility at FMR is a central quantity.

In this work, we obtain general expressions of the dynamic magnetic
susceptibility matrix of an arbitrary magnetic material near FMR based on
the casuality principle and the fact that microwave absorption at FMR
is a Lorentzian in microwave frequency at a fixed static magnetic field.
The dynamic magnetic susceptibility is not a Polder tensor in general.
It becomes a Polder tensor when the sample is isotropic or uniaxial with
the static magnetic field along its easy-axis. The matrix is fully
characterized by a few constants that can be determined by traditional
microwave absorption experiments. Interestingly, the linear response
coefficient of the magnetization to the total rf field, the sum of the
applied external rf field and the internal rf field due to the
magnetization precession, is a Polder tensor. However, this coefficient is
not the dynamic magnetic susceptibility, and cannot be measured directly.
Furthermore, under a fixed rf, the susceptibility matrix elements may
have multiple peaks in the field at a fixed microwave frequency in
contrast to a single peak in frequency at a fixed field.
The multiple field peaks come from the non-monotonic behavior of the total
effective magnetic field to the externally applied static magnetic field.
With the knowledge on the dynamic magnetic susceptibility matrix, it is
possible to separate the contributions of the AMR to the dc-voltage from
that of the AHE. We also show that another broad peak may appear in the
susceptibility matrix elements if the resonance frequency is not very
sensitive to the externally applied magnetic field.
In turn, a broad peak might also appear in the field-dependence of the
microwave absorption curves as well as the dc-voltage lineshape.
This peak is not from the resonance, and its shape is not Lorentzian.
The paper is organized as follows. In Sec. II we first reformulate the
generalized Ohm's law for a magnetic metal, and explain how a dc-voltage can
come out of the spin rectification at FMR. Then we derive the expression
of the dynamic magnetic susceptibility matrix both as a function of
microwave frequency or as a function of static magnetic field.
The experimental method of determining the matrix is then given.
With the dynamic magnetic susceptibility matrix, we analyse the dc-voltage
from AMR and AHE. Section III uses several examples to verify the general
form of the dynamic magnetic susceptibility matrix obtained in Sec. II.
We show how multiple FMR peaks in field at a fixed frequency can come from
the non-monotonic behavior of the effective field in applied field, and how
a non-resonance peak can arise when the resonance frequency (or effective
field) is not sensitive to the externally applied static magnetic field.
The conclusion is given in Sec. IV, followed by acknowledgement.

\section{Theoretical analysis}

\subsection{The generalized Ohm's law and spin rectification}
The electrical detection of FMR is based on the generalized Ohm's law
in a polycrystalline ferromagnetic metal \cite{juretschke, egan, harder}
\begin{equation}
\mathbf E=\rho_{\perp}\mathbf J+\frac{\Delta\rho}{M^2}(\mathbf J\cdot\mathbf
M)\mathbf M-R_0\mathbf J\times\mathbf H-R_1\mathbf J\times\mathbf M,
\label{ohms_law}
\end{equation}
where $M$ is the magnitude of magnetization $\mathbf M$, $\mathbf H$
and $\mathbf J$ are respectively the magnetic field and the electric
current density. $\rho_{\perp}$ is the longitudinal resistivity
when $\mathbf M$ and $\mathbf J$ are perpendicular with each other.
$\Delta\rho=\rho_{||}-\rho_{\perp}$ is the difference between $\rho_
{\perp}$ and the longitudinal resistivity $\rho_{||}$ when $\mathbf M$
is parallel to $\mathbf J$, and the $\Delta\rho$-term is called AMR.
$R_0$ and $R_1$ describe respectively the ordinary and anomalous Hall
effects.

Eq. \eqref{ohms_law} is in fact the most general linear response of electric
field of a polycrystalline magnet to an applied electric current density.
It could be understood by the following reasoning. For simplicity,
let us assume that there is no external field ($\mathbf H=0$).
For the linear response of the electric field $\mathbf E$ to an
electric current density $\mathbf J$, the most general expression
of $\mathbf E$ must be $\mathbf E=\tensor\rho(\mathbf M)\mathbf J$,
where $\tensor\rho(\mathbf M)$ is a rank-2 Cartesian tensor that
depends on $\mathbf M$ since $\mathbf M$ is the only available vector.
It is well-known that a Cartesian tensor of rank 2 can be decomposed
into the direct sum of a scalar of function of $M$, a vector that
is a function of $M$ multiplying $\mathbf M$, and a traceless symmetric
tensor of a function of $M$ multiplying $\mathbf M\mathbf M-M^2/3$.
Thus the most general expression of $\mathbf E$ is
$\mathbf E=(\rho_{\perp}+\Delta\rho/3)\mathbf J+R_1 \mathbf M\times
\mathbf J +(\Delta\rho/M^2)(\mathbf M \mathbf M-M^2/3)\cdot\mathbf J$.
This is exactly Eq. \eqref{ohms_law}, and no new physics is obtained from
this reasoning. However, things are very different if we carry out the
similar analysis for a polycrystalline magnetic film lying in the $xy$-plane.
Although the polycrystalline sample is isotropic in the film plane, $\hat z$
is an available vector and $\tensor\rho(\mathbf M,\hat z)$ should be a
function of both $\mathbf M$ and $\hat z$. Since three vectors ($\mathbf M$,
$\hat z$, and $\mathbf M\times\hat z$) and three traceless symmetric tensors
($\mathbf M \mathbf M-M^2/3$, $\mathbf M \hat z+\hat z \mathbf M-2M_z/3$, and
$\hat z\hat z-1/3$), can be constructed out of $\mathbf M$ and $\hat z$ we
have, with similar reasoning as we used early for polycrystline magnetic bulk,
$\mathbf E=(\rho_{\perp}+\Delta\rho/3+\rho_2/3) \mathbf J+(R_1 \mathbf
M+\rho_1\hat z+ R_2' \mathbf M\times\hat z)\times\mathbf J+ [(\Delta\rho
/M^2)(\mathbf M\mathbf M-M^2/3)+R_3'(\mathbf M\hat z+\hat z\mathbf M
-2M_z/3)]\cdot\mathbf J+\rho_2(\hat z\hat z-1/3)\cdot\mathbf J=
\rho_{\perp} \mathbf J-R_1 \mathbf J \times \mathbf M+ (\Delta\rho/M^2)
(\mathbf M\cdot\mathbf J)\mathbf M-\rho_1\mathbf J \times \hat z +R_2
(\hat z\cdot\mathbf J)\mathbf M+R_3(\mathbf M\cdot\mathbf J)\hat z -R_4
M_z \mathbf J + \rho_2(\hat z\cdot\mathbf J)\hat z$, where
$R_2\equiv R_2'+R_3'$, $R_3\equiv R_3'-R_2'$ and $R_4\equiv (2/3)R_2'$.
The $\rho_1$-term may be interpreted as the spin-Hall term if the sign of
$\rho_1$ for spin-up electrons is opposite to that for spin-down electrons.
It is known that spin-orbit interactions can lead to a term like this.
Obviously, $\rho_2$ contributes to the longitudinal resistivity along
$z-$direction. Interestingly, three new terms are obtained:
1) A current perpendicular to the film induces an electric field in
$\mathbf M$ direction (the $R_2$-term). 2) A current along $\mathbf M$
direction generates an electric field in $\hat z$ direction ($R_3$-term).
3) The $R_4$-term says that resistance of the film depends on $M_z$ linearly.
Of course, the coefficients of these terms depend, in principle, on $M$.
It should be pointed out that the interfacial effects are very common
in physics. \cite{lujie} Generally speaking, all new terms should exist
in polycrystalline magnetic films. Of course, their
values depend on microscopic interactions that lead to these terms.
However, these terms are not the subjects of this study although
they deserve a careful and in-depth investigation \cite{Yin}.
We shall only concentrate on AMR and AHE in Eq. \eqref{ohms_law}.

Let us come back to dc-voltage in the electrical detection of FMR in
which electric current density $\mathbf J=\mathrm{Re}(\mathbf je^{-i
(\omega t+\phi_1)})$ comes from the rf electric field only, where
$\omega$ and $\phi_1$ are respectively the microwave frequency
and the phase lag of the electric current with the electric field.
$\phi_1$ is a material parameter that depends on the complex electric
conductivity. The rf magnetic field $\mathrm{Re}(\mathbf he^{-i(\omega
t+\phi_2)})$ exerts a torque on $\mathbf M$, causing magnetization
precession around a static magnetization $\mathbf M_0$ that is
determined by the external static magnetic field and magnetic anisotropy.
$\phi_2$ is the phase difference between the rf electric field and
$\mathbf h$ inside the sample, and its value depends on the particular
experimental setup and the material parameters like dielectric constant.
For example, $\phi_2=\phi_1$ in those experiments \cite{hoffmann,buhrman}
in which the rf magnetic field is generated by the rf electric current.
However, in other experiments \cite{saitoh,azevedo,ohno} with microwave
cavities, $\phi_2$ differs from $\phi_1$. Since both $\phi_1$ and $\phi_2$
are material parameters and depend on experimental setup, we will treat
them as input parameters.

The magnetization $\mathbf M=\mathbf M_0+\mathrm{Re}(\mathbf me^{-i(\omega
t+\phi_2)})$ contains a small component precessing at frequency $\omega$.
Phase $\phi_2$ is added in our definition of $\mathbf m$ for convenience.
Using $\langle\ldots\rangle$ to denote the time average, the dc-voltage is
given by
\begin{equation}
U=\langle\mathbf E\rangle\cdot\mathbf l,
\label{dc_voltage}
\end{equation}
where $\mathbf l$ is the displacement vector between two electrode
contact points on the sample film.
Since $\langle\mathrm{Re}(\mathbf me^{-i(\omega t+\phi_2)})\rangle=0$ and
$\langle\mathbf J\rangle=0$, the dc-voltage comes from the terms in Eq.
\eqref{ohms_law} that contain $\langle\mathrm{Re}(\mathbf je^{-i(\omega t
+\phi_1)})\cdot \mathrm{Re}(\mathbf me^{-i(\omega t+\phi_2)})\rangle$ and
$\langle\mathrm{Re}(\mathbf je^{-i(\omega t+\phi_1)})\times\mathrm{Re}(\mathbf
m e^{-i(\omega t+\phi_2)})\rangle$, resulting in the spin rectification.
$\langle\mathbf E\rangle$ reads as
\begin{equation}
\begin{gathered}
\langle\mathbf E\rangle=\frac{\Delta\rho}{2M^2}\mathrm{Re}\{(\mathbf
j^\ast\cdot\mathbf me^{i(\phi_1-\phi_2)})\mathbf M_0
+(\mathbf j^\ast\cdot\mathbf M_0)\mathbf m e^{i(\phi_1-\phi_2)}\}\\
-\frac{R_0}{2}\mathrm{Re}(\mathbf j^\ast\times\mathbf h e^{i(\phi_1-\phi_2)})
-\frac{R_1}{2}\mathrm{Re}(\mathbf j^\ast\times\mathbf m e^{i(\phi_1-\phi_2)}),
\end{gathered}
\label{dc_e_field}
\end{equation}
where the first and the second terms on the right-hand side (RHS) come from the
AMR and the last two terms arise respectively from the ordinary and anomalous
Hall effects. It is convenient to define a Cartesian coordinate system
$(x,y,z)$, where $z-$axis is along $\mathbf M_0$. The ordinary Hall effect can
be neglected since it is much smaller than the AMR and AHE in a ferromagnetic
metal \cite{juretschke,pugh}. By substituting Eq. \eqref{dc_e_field} into
Eq. \eqref{dc_voltage}, the AMR and AHE contributions to dc-voltage are
\begin{equation}
\begin{aligned}
U_\mathrm{AMR}
=\frac{\Delta\rho}{2M}\mathrm{Re}\{[(\mathbf j^\ast\cdot\mathbf m)l_z
+j_z^\ast(\mathbf m\cdot\mathbf l)]e^{i(\phi_1-\phi_2)}\},
\end{aligned}
\label{dc_AMR}
\end{equation}
\begin{equation}
\begin{aligned}
U_\mathrm{AHE}
=-\frac{R_1}{2}\mathrm{Re}[(\mathbf j^\ast\times\mathbf m)\cdot\mathbf le^{i(\phi_1-\phi_2)}].
\end{aligned}
\label{dc_AHE}
\end{equation}
Once the phase difference $\phi_1-\phi_2$ between $\mathbf j$ and
$\mathbf h$ is given by experimental setup and material parameters,
the dc-voltage depends on how $\mathbf m$ responds to $\mathbf h$, or
the dynamic magnetic susceptibility near FMR.

\subsection{Universal form of dynamic magnetic susceptibility matrix}
It is important to obtain the dynamic magnetic susceptibility
$\tensor\chi$ because the dc-voltage in electrical detection of FMR
depends on $\tensor\chi$.
Under a microwave radiation, magnetization precession is governed
by the Landau-Liftshitz-Gilbert (LLG) equation\cite{gilbert}:
\begin{equation}
\frac{\partial\mathbf M}{\partial t}=-\gamma\mathbf M\times\mathbf H_\mathrm
{eff}+\frac{\alpha}{M}\mathbf M\times\frac{\partial\mathbf M}{\partial t},
\label{llg_eq}
\end{equation}
where $\mathbf H_\mathrm{eff}$ is the effective field which includes
the applied static magnetic field $\mathbf H$, rf magnetic field
$\mathrm{Re}(\mathbf h e^{-i(\omega t+\phi_2)})$ and anisotropy field.
$\alpha$ is the Gilbert damping coefficient. This is a nonlinear
equation that has many interesting physics, including spin wave
emission by magnetic domain wall motion \cite{hubin}, and can only be
solved for a few special issues \cite{zhouzhou}. However, if the
amplitude of $\mathbf h$ is small enough, the linear response of
$\mathbf m$ to $\mathbf h$ can be obtained from the linearized LLG Eq.
and takes the following form,
\begin{equation}
\left(
\begin{matrix}
m_x \\ m_y \\ m_z
\end{matrix}
\right)=\left(
\begin{matrix}
\chi_{xx} & \chi_{xy} & 0\\
\chi_{yx} & \chi_{yy} & 0\\
0 & 0 & 0
\end{matrix}
\right)\left(
\begin{matrix}
h_x \\ h_y \\ h_z
\end{matrix}
\right).
\label{chi}
\end{equation}
The structure of Eq. \eqref{chi} comes from the facts that $\mathbf m$ is
perpendicular to $\mathbf M_0$ ($m_z=0$) and $h_z$ does not exert any
torque on $\mathbf M$ so that $\mathbf m$ does not depend on $h_z$.

The energy change rate of the system can be written as
\begin{equation}
\frac{d\varepsilon}{dt}=
\nabla_\mathbf M \varepsilon\cdot\frac{\partial\mathbf M}{\partial t}
-\mu_0 \mathbf M \cdot\frac{\partial\mathbf H_\mathrm{e}}{\partial t}
\label{dE_dt}
\end{equation}
where $\varepsilon$ is the energy density of the system, and $\mathbf H_
\mathrm{e}\equiv \mathbf H +\mathrm{Re}(\mathbf h e^{-i(\omega t+\phi_2)})$
is the total applied field. Notice Eq. \eqref{llg_eq} can also be casted as
\begin{equation}
\frac{\partial\mathbf M}{\partial t}=-\frac{\gamma}{1+\alpha^2}\mathbf
M\times\mathbf H_\mathrm{eff}-\frac{\alpha\gamma}{(1+\alpha^2)M}
\mathbf M\times(\mathbf M\times\mathbf H_\mathrm{eff}).
\label{llg_eq2}
\end{equation}
Substituting $\nabla_\mathbf M \varepsilon=-\mu_0\mathbf H_\mathrm{eff}$
and Eq. \eqref{llg_eq2} into Eq. \eqref{dE_dt}, one obtains
\begin{equation}
\frac{d\varepsilon}{dt}=
-\frac{\alpha\gamma\mu_0}{(1+\alpha^2)M}
|\mathbf M\times\mathbf H_\mathrm{eff}|^2
-\mu_0\mathbf M\cdot\dot{\mathbf H}_\mathrm{e}.
\label{dE_dt2}
\end{equation}
The first term on the RHS of Eq. \eqref{dE_dt2} is the energy dissipation
due to the motion of magnetization while the second term describes the
energy release/absorption rate from the microwave. Since the total energy
of a system cannot increase or decrease indefinitely, the time average of
the energy change rate should be zero, $\langle\dot{\varepsilon}\rangle=0$
so that the average microwave absorption (by the sample) rate $I$
should be equal to energy dissipation rate \cite{zzsun}. $I$ reads
\begin{equation}
I=-\mu_0\langle\mathbf M\cdot\dot{\mathbf H}_\mathrm{e}\rangle
=-\frac{\mu_0\omega}{2}\mathrm{Im}(\mathbf{m}^\ast\cdot\mathbf{h}),
\label{absorp}
\end{equation}
and by substituting Eq. \eqref{chi} into Eq. \eqref{absorp}, $I$,
in terms of $\tensor\chi$, becomes
\begin{equation}
\begin{aligned}
I=\frac{\mu_0\omega}{2}[h_x^\ast h_x\chi_{xx}^{\prime\prime}
+\mathrm{Re}(h_x^\ast h_y)(\chi_{xy}^{\prime\prime}+\chi_{yx}^{\prime\prime})\\
+\mathrm{Im}(h_x^\ast h_y)(\chi_{xy}^\prime-\chi_{yx}^\prime)
+h_y^\ast h_y\chi_{yy}^{\prime\prime}],
\end{aligned}
\label{absorp_chi}
\end{equation}
where $\chi_{jk}=\chi'_{jk}+i\chi''_{jk}$ ($j,k=x,y$).
It is known that microwave absorption intensity as a function of
microwave frequency at a fixed static magnetic field is Lorentzian.
We shall use this property and the causality principle (events come
after causes) \cite{toll} to find both $\tensor\chi(\omega)$ and
$\tensor\chi(H)$ below.

\subsubsection{Frequency dependence of $\tensor\chi$}
In the case that the microwave frequency $\omega$ is swept at a fixed
static magnetic field $H$, each element of $\tensor\chi(\omega)$ should be
analytic in the up-half complex $\omega$-plane due to the causality
principle (events come after causes)\cite{toll}. The consequence is that
its real and imaginary parts are related to each other by the Kramers-Kronig
relations. In the resonance region and in the absence of high-order
contribution, the energy absorption intensity is a Lorentzian function
$L(\omega,\omega_0,\Gamma)$,
\begin{equation}
L(\omega, \omega_0,\Gamma)=
\frac{1}{\pi}\frac{\frac{\Gamma}{2}}{(\omega-\omega_0)^2+(\frac{\Gamma}{2})^2}.
\label{lorentz}
\end{equation}
The counterpart of $L(\omega,\omega_0,\Gamma)$ in the Kramers-Kronig
relations is the function $D(\omega,\omega_0,\Gamma)$,
\begin{equation}
D(\omega,\omega_0,\Gamma)=
\frac{1}{\pi}\frac{\omega-\omega_0}{(\omega-\omega_0)^2+(\frac{\Gamma}{2})^2},
\label{disp}
\end{equation}
where $\omega_0$ denotes the resonance frequency and $\Gamma$ is the
linewidth which is a positive number. $D$ can be both positive and negative,
and extends further in comparison to $L(\omega,\omega_0,\Gamma)$.
Since the energy absorption lineshape should be Lorentzian for an
arbitrary $\mathbf h$, each term on the RHS of Eq. \eqref{absorp_chi}
should have the form of Eq. \eqref{lorentz} with the same resonance
frequency $\omega_0$ and linewidth $\Gamma$:
\begin{equation}
\begin{aligned}
\chi_{xx}^{\prime\prime}(\omega) &=
C_1(\frac{\pi\Gamma}{2})L(\omega,\omega_0,\Gamma),\\
\chi_{xy}^{\prime\prime}(\omega)+\chi_{yx}^{\prime\prime}(\omega) &=
2C_2(\frac{\pi\Gamma}{2})L(\omega,\omega_0,\Gamma),\\
\chi_{xy}^\prime(\omega)-\chi_{yx}^\prime(\omega) &=
2C_3(\frac{\pi\Gamma}{2})L(\omega,\omega_0,\Gamma),\\
\chi_{yy}^{\prime\prime}(\omega) &=
C_4(\frac{\pi\Gamma}{2})L(\omega,\omega_0,\Gamma),\\
\end{aligned}
\label{chi_f_l}
\end{equation}
where the four {\it real} numbers $C_1\sim C_4$ are determined by the
material parameters and experimental setup. A factor of $\pi\Gamma/2$
is introduced in order to make $C_i\ (i=1,2,3,4)$ dimensionless.
By using the Kramers-Kronig relations, one can write
\begin{equation}
\begin{aligned}
\chi_{xx}^\prime(\omega) &=
-C_1(\frac{\pi\Gamma}{2})D(\omega,\omega_0,\Gamma),\\
\chi_{xy}^\prime(\omega)+\chi_{yx}^\prime(\omega) &=
-2C_2(\frac{\pi\Gamma}{2})D(\omega,\omega_0,\Gamma),\\
\chi_{xy}^{\prime\prime}(\omega)-\chi_{yx}^{\prime\prime}(\omega) &=
2C_3(\frac{\pi\Gamma}{2})D(\omega,\omega_0,\Gamma),\\
\chi_{yy}^\prime(\omega) &=
-C_4(\frac{\pi\Gamma}{2})D(\omega,\omega_0,\Gamma).\\
\end{aligned}
\label{chi_f_d}
\end{equation}
From Eqs. \eqref{chi_f_l} and \eqref{chi_f_d}, $\tensor\chi(\omega)$
should have the general form of
\begin{equation}
\tensor\chi(\omega)=
\frac{\pi\Gamma}{2}[L(\omega,\omega_0,\Gamma)+iD(\omega,\omega_0,\Gamma)]\tensor C,
\label{chi_f}
\end{equation}
where
\begin{equation}
\tensor C=\left(
\begin{matrix}
iC_1 & C_3+iC_2 & 0\\
-C_3+iC_2 & iC_4 & 0\\
0 & 0 & 0\\
\end{matrix}
\right).
\label{C}
\end{equation}
Thus $\tensor\chi(\omega)$ of an arbitrary magnet is completely determined
by six numbers $\omega_0$, $\Gamma$, $C_1$, $C_2$, $C_3$ and $C_4$.
Interestingly, the two off-diagonal matrix elements of $\tensor\chi(\omega)$
are not opposite to each other in general, because $C_2$ is zero only when
the material has certain special symmetry or for a special coordinate
(see discussion in Section II.C).

In terms of phase $\psi\equiv\cot^{-1}[-D/L]=\cot^{-1}[(\omega_0-\omega)/\Gamma]$,
$\mathbf m$ relates to $\mathbf h$ as
\begin{equation}
\mathbf m=\left(
\begin{matrix}
|\chi_{xx}| & |\chi_{xy}|e^{-i\cot^{-1}\frac{C_2}{C_3}} & 0\\
|\chi_{yx}|e^{i\cot^{-1}\frac{C_2}{C_3}} & |\chi_{yy}| & 0\\
0 & 0 & 0
\end{matrix}
\right)\mathbf h e^{i\psi}.
\label{phase_lag}
\end{equation}
When microwave frequency $\omega$ is swept through FMR region at a fixed
static magnetic field, the phase $\psi$ changes from $0^\mathrm o$ to
$180^\mathrm o$ and is $90^\mathrm o$ at FMR. This property of $\psi$
is universal for an arbitrary system.

\subsubsection{Field dependence of $\tensor\chi$}
In most experiments, the microwave frequency was fixed and the applied
static magnetic field was swept. Thus, it is also useful to obtain the
field dependence of $\tensor\chi$. One cannot directly apply the
causality principle here, and we shall use Eq. \eqref{chi_f}.
Notice that each of $\omega_0$, $\Gamma$ and $\tensor C$ is a function
of $H$, $\tensor\chi$ can be expressed as
\begin{equation}
\begin{aligned}
\tensor\chi(H) &= \frac{\pi\Gamma(H)}{2}[L(\omega,\omega_0(H),\Gamma(H))\\
&+iD(\omega,\omega_0(H),\Gamma(H))]\tensor C(H).
\end{aligned}
\label{chi_H}
\end{equation}
Unlike the frequency sweeping case where $\mathbf M_0$ is always in the
same direction, $\mathbf M_0$ changes its direction during the field
sweeping process (in which the direction of $\mathbf H$ does not change).
However, one expects that $\omega_0$, $\Gamma$ and $\tensor C$ do not
change much if the FMR peak width is narrow.
Thus one can replace $\mathbf M_0$, $\Gamma$ and $\tensor C$ by
their values at the resonance field $H_0$ where $\omega_0(H_0)=\omega$,
\begin{equation}
\begin{aligned}
\mathbf M_0(H) &\approx \mathbf M_0(H_0),\\
\Gamma(H) &\approx \Gamma(H_0),\\
\tensor C(H) &\approx \tensor C(H_0).
\end{aligned}
\label{approx}
\end{equation}
Similarly, one can expand $\omega_0(H)$ around $H_0$ to the linear order
in $H-H_0$,
\begin{equation}
\omega_0(H)\approx\omega+\beta(H-H_0).
\label{f-H}
\end{equation}
Substituting Eqs. \eqref{approx} and \eqref{f-H} into Eq. \eqref{chi_H},
one has
\begin{equation}
\tensor\chi(H)=\frac{\pi\Gamma_1}{2}[L(H,H_0,\Gamma_1)
+i\frac{-\beta}{|\beta|}D(H,H_0,\Gamma_1)]\tensor C(H_0),
\label{chi_H2}
\end{equation}
where $\Gamma_1$ is the linewidth of the field:
\begin{equation}
\Gamma_1=\frac{\Gamma(H_0)}{|\beta|}.
\label{widthes}
\end{equation}
It is not surprising to see that $\Gamma_1$ is large if $\omega_0(H)$
does not change much around $H_0$ (small $\beta$) while it is smaller for
a larger $\beta$. Similar to the frequency dependence of $\tensor\chi$,
the field dependence of $\tensor\chi$ around a FMR is fully determined
by six numbers, $H_0$, $\Gamma_1$, $C_1$, $C_2$, $C_3$ and $C_4$.
However, the sign in front of $D$ function in Eq. \eqref{chi_H2} is
decided by the sign of
$\beta$ (``$+$" when  $\omega_0$ decreases with $H$ and ``$-$" otherwise).
In terms of phase $\psi\equiv\cot^{-1}[(\beta/|\beta|)D/L]
=\cot^{-1}[(\beta/|\beta|)(H-H_0)/\Gamma_1]$, the $\beta$-sign decides
whether $\psi$ changes from $180^\mathrm o$ to $0^\mathrm o$
or from $0^\mathrm o$ to $180^\mathrm o$
when the field is swept up through the resonance region.
The sign also relates to the sign of dc voltage
as well as whether the observed signal, like $dI/dH$ in traditional
microwave absorption measurement, increases or decreases with $H$.
This is in contrast to the frequency dependence of $\tensor\chi$ as shown
in Eq. \eqref{chi_f}, where the sign before $D$ function is always
positive so that phase $\psi$ always increases from $0^\mathrm o$ far under
the resonance frequency ($\omega \ll \omega_0$) to $180^\mathrm o$ far
above the resonance frequency ($\omega \gg\omega_0$). Obviously,
$\tensor\chi(H)$ is not a Polder tensor in general. $\tensor\chi(H)$ may
deviate from Eq. \eqref{chi_H2} when approximation of Eq. \eqref{approx}
is not good. As a result, a non-resonance broad peak and symmetry changes
in elements of $\tensor\chi(H)$ can occur. Explicit examples will
be discussed in Section III.B.

\subsection{The Polder or Non-Polder tensors}

In order to understand why the Polder tensor was always used in experimental
analyses, we want to see which linear response coefficient is a Polder tensor.
Assume $\mathbf H_\mathrm K(\mathbf M)$ is the anisotropy field,
then, to the first order in $\mathbf m$, the effective field is
\begin{equation}
\begin{gathered}
\mathbf H_\mathrm{eff}=\mathbf H+\mathbf H_\mathrm K(\mathbf M_0)+\\
\nabla_\mathbf M\mathbf H_\mathrm K|_{\mathbf M_0}\cdot\mathbf
m e^{-i(\omega t+\phi_2)} +\mathbf h e^{-i(\omega t+\phi_2)}.
\end{gathered}
\label{h_eff}
\end{equation}
Substituting Eq. \eqref{h_eff} and
$\mathbf M=\mathbf M_0+\mathbf m e^{-i(\omega t+\phi_2)}$ into
Eq. \eqref{llg_eq}, one has, up to the linear term in $\mathbf m$,
\begin{equation}
-i\omega\mathbf m=
-\gamma\mathbf m\times\mathbf H_\mathrm{dc}
-\gamma\mathbf M_0\times\mathbf h_\mathrm{ac}
-i\omega\frac{\alpha}{M}\mathbf M_0\times\mathbf m,
\label{llg_eq3}
\end{equation}
where $\mathbf H_\mathrm{dc}$ is the effective dc-field
\begin{equation}
\begin{gathered}
\mathbf H_\mathrm{dc}=\mathbf H+\mathbf H_\mathrm K(\mathbf M_0)
\end{gathered}
\label{h_dc}
\end{equation}
and $\mathbf h_\mathrm{ac}$ is the effective ac-field that is the sum
of applied rf field $\mathbf h$ and internal rf field due to the
precessing magnetization,
\begin{equation}
\begin{gathered}
\mathbf h_\mathrm{ac}=\mathbf h
+\nabla_\mathbf M\mathbf H_\mathrm K|_{\mathbf M_0}\cdot\mathbf m.
\end{gathered}
\label{h_ac}
\end{equation}
Obviously, the solution of Eq. \eqref{llg_eq3} is
\begin{equation}
\mathbf m=\left(
\begin{matrix}
\chi_L & -i\chi_T & 0\\
i\chi_T & \chi_L & 0\\
0 & 0 & 0\\
\end{matrix}
\right)
\mathbf h_\mathrm{ac}=\tensor\chi_\mathrm P\cdot \mathbf h_\mathrm{ac},
\label{Polder}
\end{equation}
where
\begin{equation}
\begin{aligned}
\chi_L=\frac{\omega_r\omega_M}{\omega_r^2-\omega^2},
\chi_T=\frac{\omega\omega_M}{\omega_r^2-\omega^2}
\end{aligned}
\end{equation}
with $\omega_M=\gamma M$ and $\omega_r=\gamma H_\mathrm{dc}-i\alpha\omega$.
$\tensor\chi_\mathrm P$ is the so-called Polder tensor \cite{seidel} and is
often assumed to be the dynamic magnetic susceptibility in analyses.
The Polder tensor $\tensor\chi_\mathrm P$
is the linear response coefficient of the magnetization to the effective
ac-field $\mathbf m=\tensor\chi_\mathrm P \mathbf h_\mathrm{ac}$.
Different from $\tensor\chi$ which is the linear response coefficient
of $\mathbf m$ to the applied rf field $\mathbf m=\tensor\chi\mathbf h$
and measurable, $\tensor\chi_\mathrm P$ {\it is not measurable because}
$\mathbf h_\mathrm{ac}$ contains both the applied rf field $\mathbf h$
and an oscillating anisotropy field due to the precessing magnetization.
Obviously $\mathbf h_\mathrm{ac}$ {\it cannot be measured directly}.
If one defines matrix $\tensor W$ as $\tensor W\equiv (1+\tensor\chi_
\mathrm P\cdot\frac{\delta^2 \varepsilon}{\mu_0 \delta \mathrm M \delta
\mathrm M}|_{\mathrm M_0})^{-1}$, then $\tensor\chi$ relates to $\tensor
\chi_\mathrm P$ as $\tensor\chi=\tensor W\cdot \tensor\chi_\mathrm P$.
$\tensor\chi_\mathrm P$ is simple and can be fully determined by a few
constants even though it is not a true measurable quantity while $\tensor
\chi$ is much more complicated and is sensitive to the magnetic anisotropy.
This may be the reason why $\tensor\chi_\mathrm P$ was so often taken as
the dynamic magnetic susceptibility matrix although the real one is
$\tensor\chi$.

\begin{figure}
\centering
\includegraphics[width=0.4\textwidth]{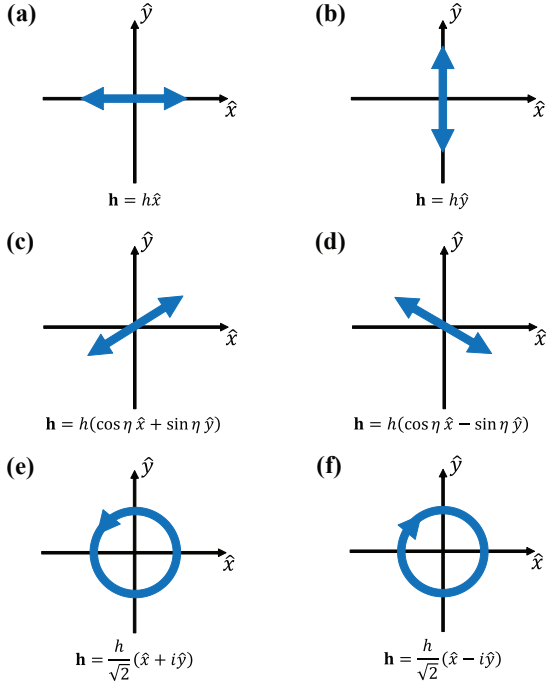}\\
\caption{Different polarizations of the applied microwave
magnetic field $\mathbf h$.
$\mathbf h$ is linearly polarized in (a)-(d) and circularly
polarized in (e) and (f). The $z-$axis is along $\mathbf M_0$
and $0<\eta<\pi/2$ (in (c) and (d)).
}
\label{recipe}
\end{figure}

It should be interesting to know when $\tensor\chi$ is a Polder tensor,
i.e. $\chi_{xy}=-\chi_{yx}$ and $\chi_{xx}=\chi_{yy}$ or $C_2=0$ and
$C_1=C_4$, according to Eq. \eqref{C}.
Let us first understand the physical meanings of $C_1\sim C_4$.
Consider six different $\mathbf h$ of amplitude $h$ shown in Fig.
\ref{recipe}: along the $x-$direction $\mathbf h=h\hat x$ (a); the
$y-$direction $\mathbf h=h\hat y$ (b); the direction $\eta$ angle with
$x-$axis $\mathbf h=h(\cos\eta \hat x+\sin\eta \hat y)$ (c); the direction
$\pi-\eta$ angle with the $x-$axis $\mathbf h=h(\cos\eta \hat x -\sin\eta
\hat y)$ (d); the right-hand circularly polarized field
$\mathbf h=\frac{h}{\sqrt{2}}(\hat x+i\hat y)$ (e); the left-hand circularly
polarized field $\mathbf h=\frac{h}{\sqrt{2}}(\hat x-i\hat y)$ (f).
If one denotes the microwave absorption intensities around FMR in Figs.
\ref{recipe}(a)$\sim$\ref{recipe}(f) by $I_a\sim I_f$, from Eqs.
\eqref{absorp_chi}, \eqref{chi_f_l} and \eqref{chi_f_d}, one has
$I_a=(0.25h^2\mu_0\omega\pi\Gamma)C_1 L(\omega,\omega_0,\Gamma)$ and
$I_b=(0.25h^2\mu_0\omega\pi\Gamma)C_4 L(\omega,\omega_0,\Gamma)$,
proportional to $C_1$ and $C_4$ respectively. Similarly,
$I_c-I_d=(0.5h^2\mu_0\omega\pi\Gamma)\sin(2\eta)C_2 L(\omega,\omega_0,\Gamma)$
is proportional to $C_2$ and $I_e-I_f=(h^2\mu_0\omega\pi\Gamma)C_3 L(\omega,
\omega_0,\Gamma)$ measures $C_3$. Obviously, $C_1$ and $C_4$ are always
positive whereas $C_2$ and $C_3$ can be positive, negative or zero,
depending on the experimental setup and the choice of coordinate.
Thus, one has $\chi_{xy}=-\chi_{yx}$ if $C_2$ is zero (or $I_c=I_d$),
and vice versa.

The condition for $\chi_{xy}=-\chi_{yx}$ can be obtained from the
exact solution of $\tensor\chi$ for an arbitrary energy landscape
$\varepsilon(\mathbf M)$. By solving Eq. \eqref{llg_eq3},
$\tensor\chi$ is
\begin{equation}
\tensor\chi=\frac{1}{|T|^2-PQ}
\left(
\begin{matrix}
P & T & 0 \\
T^\ast & Q & 0\\
0 & 0 & 0
\end{matrix}
\right),
\label{chi_PQT}
\end{equation}
with
\begin{equation}
\begin{aligned}
P &= -\frac{1}{\mu_0}\frac{\partial^2\varepsilon}{\partial M_y^2}\big|_{\mathbf M_0}
-\frac{H_\mathrm{dc}}{M}+i\alpha\frac{\omega}{\gamma M},\\
Q &= -\frac{1}{\mu_0}\frac{\partial^2\varepsilon}{\partial M_x^2}\big|_{\mathbf M_0}
-\frac{H_\mathrm{dc}}{M}+i\alpha\frac{\omega}{\gamma M},\\
T &= \frac{1}{\mu_0}\frac{\partial^2\varepsilon}{\partial M_x\partial M_y}\big|_{\mathbf M_0}
+i\frac{\omega}{\gamma M}.\\
\end{aligned}
\label{PQT}
\end{equation}
Thus, one can have $\chi_{xy}=-\chi_{yx}$ only when
\begin{equation}
\begin{aligned}
\frac{\partial^2\varepsilon}{\partial M_x\partial M_y}\big|_{\mathbf M_0}=0.
\end{aligned}
\label{st}
\end{equation}
This could happen in several cases. If the system has rotation symmetry
about $\mathbf M_0$ so that the energy absorption rate of a linearly
polarized microwave does not depend on the direction of $\mathbf h$ in the
$xy-$plane, then one has $C_2\propto I_c-I_d=0$ and $C_1/C_4=I_a/I_b=1$.
Thus, $\chi_{xy}=-\chi_{yx}$ and $\chi_{xx}=\chi_{yy}$, and $\tensor\chi$
is a Polder tensor. An obvious example for this case is an isotropic system
without magnetic anisotropy so that $\mathbf M_0$ is along $\mathbf H$.
$\chi_{xy}=-\chi_{yx}$ if the free energy density is symmetric about the
$yz-$ or $xz-$plane, i.e. $\varepsilon(M_x,M_y,M_z)=\varepsilon(-M_x,M_y,M_z)$
or $\varepsilon(M_x,M_y,M_z)=\varepsilon(M_x,-M_y,M_z)$.
Take derivatives with respect to $M_x$ and $M_y$ on both sides, one has
\begin{equation}
\frac{\partial^2\varepsilon}{\partial M_x\partial M_y}\big|_{\mathbf M_0}=
-\frac{\partial^2\varepsilon}{\partial M_x\partial M_y}\big|_{\mathbf M_0}=0.
\nonumber
\end{equation}

For an arbitrary magnetic material, it is also possible to make
$\chi_{xy}=-\chi_{yx}$ by properly choosing $x$ and $y$ axes.
To prove this statement, let $\tensor\chi$, characterized by $C_1\sim C_4$,
be a non-Polder tensor in the original $xy-$coordinate. One can choose
a new $x'y'-$coordinate that relates to $xy-$coordinate by the rotation
matrix (around the $z-$axis)
\begin{equation}
\tensor R=\left(
\begin{matrix}
\cos\delta & -\sin\delta & 0\\
\sin\delta & \cos\delta & 0\\
0 & 0 & 1\\
\end{matrix}
\right).
\label{rotation}
\end{equation}
In this new coordinate, the susceptibility matrix is
$\tensor R\tensor\chi\tensor R^T$ characterized by new $C_1'\sim C_4'$.
$C_1'\sim C_4'$ relate to $C_1\sim C_4$ by
\begin{equation}
\begin{aligned}
C_1' &= C_1\cos^2\delta-C_2\sin2\delta+C_4\sin^2\delta,\\
C_2' &= \frac{C_1-C_4}{2}\sin2\delta+C_2\cos2\delta,\\
C_3' &= C_3,\\
C_4' &= C_1\sin^2\delta+C_2\sin2\delta+C_4\cos^2\delta.\\
\end{aligned}
\label{C_prime}
\end{equation}
If one chooses $\delta=0.5\arctan\frac{2C_2}{C_4-C_1}$, $C_2'=0$ so that
off-diagonal elements are opposite to each other in the new coordinate.
Interestingly, $C_3$ does not change while $C_1$, $C_2$ and $C_4$
depend on the coordinate change.
This is not surprising because $C_3$ relates to the microwave absorption
rate difference between the right-hand and left-hand circularly
polarized rf fields which rotate around $z-$axis so that the choice
of $x-$ and $y-$axes is not matter to the energy absorption, or to $C_3$.

Naively, one may expect that the traditional microwave absorption
experiments cannot measure a complex number such as susceptibility
matrix elements, because only Lorentzian lineshapes can be measured.
However, the above discussions clearly show that the traditional
microwave absorption experiments are capable of completely determining
the dynamic magnetic susceptibility matrix.

\subsection{Determination of matrix $\tensor \chi$}

One of the achievements in subsection II.B is that the dynamic magnetic
susceptibility can be fully characterized by a few parameters.
This is important because $\tensor \chi$ expressed in terms of magnetic
anisotropy, like Eqs. (31) and (32), is not too useful since the exact form of
the magnetic anisotropy is hard to obtain in reality if not impossible.
A natural question is how to experimentally determine these parameters.
From the early discussion of the meaning of $C_1\sim C_4$, a recipe, which
is based on traditional FMR measurements and allows a full determination of
these parameters can be obtained.
The idea is to apply different $\mathbf h$ as shown in Fig. \ref{recipe}
so that one can associate $C_1\sim C_4$, $\omega_0$ (or $H_0$) and
$\Gamma$ (or $\Gamma_1$) with peak-height, peak-position, and peak-width
of microwave absorption spectrum. Our recipe is:

\emph{Step I} Use usual magnetic measurement to locate the direction of
$\mathbf M_0$ along which the $z-$axis is chosen, and a $xyz-$Cartesian
coordinate can be assigned.

\emph{Step II} Conduct a usual FMR microwave absorption experiment by
using a microwave whose rf magnetic field $\mathbf h$ is along $x-$axis.
Either frequency $\omega$ or field $H$ is swept so that resonance $\omega_0$
(or $H_0$), together with linewidth $\Gamma$ (or $\Gamma_1$) can be obtained
from the microwave absorption curve.
As discussed early (Fig. \ref{recipe}), $C_1$ is given by the peak-height $I_a$
of the microwave absorption curve as $C_1=2I_a/(\mu_0h^2\omega_0)$.

\emph{Step III} Repeat Step II but use $\mathbf h=h \hat y$, $C_4$ is
given by the peak-height $I_b$ as $C_4=2I_b/(\mu_0h^2\omega_0)$.

\emph{Step IV} Repeat Step II but use $\mathbf h=h(\cos\eta\hat x+\sin
\eta\hat y)$ first to obtain the peak-height $I_c$ of the microwave
absorption curve, then use $\mathbf h=h(\cos\eta\hat x-\sin\eta\hat y)$
to obtain the peak-height $I_d$. $C_2$ is then given by
$C_2=(I_c-I_d)/[(\mu_0h^2\omega_0)\sin2\eta]$.

\emph{Step V} Repeat Step II but use $\mathbf h=h(\cos\eta\hat x+
i\sin\eta\hat y)$ first to obtain the peak-height $I_e$ of the microwave
absorption curve, then use $\mathbf h=h(\cos\eta\hat x-i\sin\eta\hat y)$
to obtain the peak-height $I_f$. $C_3$ is then given by
$C_3=(I_e-I_f)/(2\mu_0h^2\omega_0)$.

According to Eq. \eqref{chi_f}, $\tensor\chi(\omega)$ is fully determined
from the above steps. For $\tensor\chi(H)$, one needs to determine whether
$\beta/|\beta|=-1$ or $\beta/|\beta|=1$, corresponding to the increase or
decreases of $\omega_0$ with $H$. Step V involves the circularly-polarized
microwaves that may be difficult to obtain in experiments.
However, one should notice that the recipe is not unique.
One may use other easily obtained microwaves to determine $C_3$.

\subsection{The lineshape of dc-voltage}
With the general form of the dynamic magnetic susceptibility matrix
$\tensor\chi$, we can compute the AMR and AHE contributions to the dc-voltage
and study the symmetry of dc-voltage lineshapes.
Substituting Eq. \eqref{chi} into Eqs. \eqref{dc_AMR} and \eqref{dc_AHE},
$U_\mathrm{AMR}$ and $U_\mathrm{AHE}$ in terms of $\tensor\chi$ are
\begin{equation}
U_\mathrm{AMR}
=-\frac{\Delta\rho}{2M}\mathrm{Re}[(j_i^\ast h_jl_z\chi_{ij}
+j_z^\ast h_jl_i\chi_{ij})e^{i(\phi_1-\phi_2)}],
\label{dc_AMR_chi}
\end{equation}
\begin{equation}
U_\mathrm{AHE}=
\frac{R_1}{2}\mathrm{Re}(\epsilon_{ijk}j_j^\ast h_ll_i\chi_{kl}e^{i(\phi_1-\phi_2)}),
\label{dc_AHE_chi}
\end{equation}
where subscript indices $i$, $j$, $k$ and $l$ can be $x$, $y$ and $z$.
Throughout this paper, $\epsilon_{ijk}$ is the Levi-Civita symbol,
and the Einstein summation convention is used.
Whether a matrix element of $\tensor\chi$ is involved in dc-voltage
depends on the applied microwave fields and experimental setup.
Substituting Eqs. \eqref{chi_f} and \eqref{C} into Eqs. \eqref{dc_AMR_chi}
and \eqref{dc_AHE_chi}, the frequency dependence of dc-voltage can be
expressed in terms of Lorentzian and $D$ functions,
\begin{equation}
U_\mathrm{AMR}(\omega)=A_1\frac{\pi\Gamma}{2}L(\omega,\omega_0,\Gamma)
+A_2\frac{\pi\Gamma}{2}D(\omega,\omega_0,\Gamma),
\label{dc_AMR_f}
\end{equation}
\begin{equation}
U_\mathrm{AHE}(\omega)=A_3\frac{\pi\Gamma}{2}L(\omega,\omega_0,\Gamma)
+A_4\frac{\pi\Gamma}{2}D(\omega,\omega_0,\Gamma),
\label{dc_AHE_f}
\end{equation}
where the relative intensities $A_1\sim A_4$ are
\begin{equation}
\begin{aligned}
A_1 &= -\frac{\Delta\rho}{2M}\mathrm{Re}[(j_i^\ast h_j l_z C_{ij}
+j_z^\ast h_j l_i C_{ij})e^{i(\phi_1-\phi_2)}],\\
A_2 &= \frac{\Delta\rho}{2M}\mathrm{Im}[(j_i^\ast h_j l_z C_{ij}
+j_z^\ast h_j l_i C_{ij})e^{i(\phi_1-\phi_2)}],\\
A_3 &= \frac{R_1}{2}\mathrm{Re}(\epsilon_{ijk}j_j^\ast h_l l_i C_{kl}e^{i(\phi_1-\phi_2)}),\\
A_4 &= -\frac{R_1}{2}\mathrm{Im}(\epsilon_{ijk}j_j^\ast h_l l_i C_{kl}e^{i(\phi_1-\phi_2)}), \\
\end{aligned}
\label{A}
\end{equation}
where subscript indices $i$, $j$, $k$ and $l$ are $x$, $y$ and $z$.
$C_{ij}$ is the element of the $i$'th row and the $j'$th column of
matrix $\tensor C$ defined in Eq. \eqref{C}.
The way to find the field dependence of dc-voltage is similar as that for
$\tensor\chi(H)$. $A_1\sim A_4$ are functions of $H$, therefore, from Eqs.
\eqref{dc_AMR_f} and \eqref{dc_AHE_f}, one has
\begin{equation}
\begin{aligned}
U_\mathrm{AMR}(H) &=
A_1(H)\frac{\pi\Gamma(H)}{2}L(\omega,\omega_0(H),\Gamma(H))\\
&+A_2(H)\frac{\pi\Gamma(H)}{2}D(\omega,\omega_0(H),\Gamma(H)),
\end{aligned}
\label{dc_AMR_H}
\end{equation}
\begin{equation}
\begin{aligned}
U_\mathrm{AHE}(H) &=
A_3(H)\frac{\pi\Gamma(H)}{2}L(\omega,\omega_0(H),\Gamma(H))\\
&+A_4(H)\frac{\pi\Gamma(H)}{2}D(\omega,\omega_0(H),\Gamma(H)).
\end{aligned}
\label{dc_AHE_H}
\end{equation}
The above expressions do not mean that $U_\mathrm{AMR}(H)$
and $U_\mathrm{AHE}(H)$ are linear combinations of Lorentzian and $D$
functions because $A_i$'s ($i=1, \ldots,4$) are also functions of $H$.
However, if the approximations of Eq. \eqref{approx} are applicable, i.e.
\begin{equation}
A_i(H)\approx A_i(H_0)\quad(i=1,2,3,4),
\label{approx_A}
\end{equation}
then $U_\mathrm{AMR}(H)$ and $U_\mathrm{AHE}(H)$ are linear combinations of
the Lorentzian and $D$ functions,
\begin{equation}
\begin{aligned}
U_\mathrm{AMR}(H) &= A_1(H_0)\frac{\pi\Gamma_1}{2}L(H,H_0,\Gamma_1)\\
&-\frac{\beta}{|\beta|}A_2(H_0)\frac{\pi\Gamma_1}{2}D(H,H_0,\Gamma_1),
\end{aligned}
\label{dc_AMR_H2}
\end{equation}
\begin{equation}
\begin{aligned}
U_\mathrm{AHE}(H) &= A_3(H_0)\frac{\pi\Gamma_1}{2}L(H,H_0,\Gamma_1)\\
&-\frac{\beta}{|\beta|}A_4(H_0)\frac{\pi\Gamma_1}{2}D(H,H_0,\Gamma_1).
\end{aligned}
\label{dc_AHE_H2}
\end{equation}
The symmetric dc-voltage lineshapes are from the Lorentzian terms, and
antisymmetric components are determined by $D$ terms.
$A_1\sim A_4$ are linear combinations of $C_1\sim C_4$ whose coefficients
depend on magnetic anisotropy and experimental setup, and their values
determine the relative weights of the symmetric and antisymmetric components.

According to Eq. \eqref{A}, the phase difference $\phi_1-\phi_2$
between rf current density and rf magnetic field is important to dc-voltage.
Several groups have found experimental evidence that the phase difference
relates to many factors like experimental setup, material parameters,
microwave frequency, etc\cite{azevedo,harder}. Therefore, $\phi_1$ and
$\phi_2$ are taken as input parameters in the following discussions.

\section{Verification and Discussion}

In this section, we use a biaxial model to verify the universal expressions
of the dynamic magnetic susceptibility matrix and dc-voltage lineshape
obtained in the last section. Both $\tensor\chi(\omega)$ and $\tensor\chi(H)$
are compared with the exact expression obtained from LLG equation \eqref{llg_eq},
as well as with numerical simulations of the LLG equation. \cite{daquino}
The universal expressions of dc-voltage due to the AMR and AHE effects near
FMR are compared with the exact numerical results based on the LLG
equation and the generalized Ohm's law of Eq. \eqref{ohms_law}.

\begin{figure}
\centering
\includegraphics[width=0.4\textwidth]{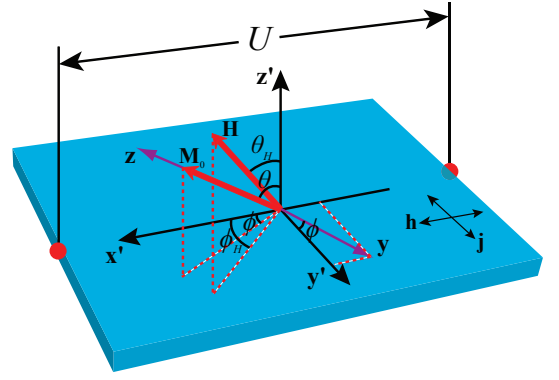}\\
\caption{Model system that mimics experimental setups in the electrical
detection of FMR. The $x'y'z'$-coordinate is fixed with respect to the sample.
The biaxial magnetic film lies in $x'y'-$ plane. The easy axis is along
$x'-$direction, and $z'-$axis is the hard axis. The $xyz$ is a moving
coordinate with the $z-$axis along $\mathbf M_0$, and the $y-$axis in the
$x'y'-$plane. $\theta$ and $\phi$ are polar and azimuthal angles of
$\mathbf M_0$ in the $x'y'z'$-coordinate, i.e. $\theta$ is the angle between
the $z-$ and $z'-$axes, and $\phi$ is the angle between in-plane component
of $\mathbf M_0$ and the $x'-$axis. $\theta_H$ and $\phi_H$ are polar and
azimuthal angles of external static field $\mathbf H$ in the $x'y'z'$-coordinate.
The directions of $\mathbf h$ and $\mathbf j$ are along the $x'-$ and
$y'-$ axes, respectively.
}
\label{configs}
\end{figure}

Our model system, which mimics popular experimental setups, is shown in
Fig. \ref{configs}. A biaxial magnetic
film lies in the $x'y'-$plane with length $l$ along the $x'-$direction.
The $x'-$ and $z'-$axes are respectively the easy and hard axes of the film.
For the simplicity, the rf magnetic field $\mathbf h$ and the rf electric
field (or rf electric current $\mathbf j$) are along the $x'-$ and $y'-$axes,
respectively. A moving $xyz-$coordinate is defined as follows.
The $z-$axis is along $\mathbf M_0$ and the $y-$axis is in the $x'y'-$plane.
$\theta$ and $\phi$ are polar and azimuthal angles of $\mathbf M_0$ in the
$x'y'z'$-coordinate while $\theta_H$ and $\phi_H$ are polar and azimuthal
angles of external static field $\mathbf H$ in the $x'y'z'$-coordinate.
Therefore, once $\mathbf M_0$ is determined, the $z-$axis is $\hat z=\sin
\theta\cos\phi\hat x'+\sin\theta\sin\phi\hat y'+\cos\theta\hat z'$, and
the $x-$ and $y-$axes are determined by $\hat x=\cos\theta\cos\phi\hat x'+
\cos\theta\sin\phi \hat y'-\sin\theta \hat z'$
and $\hat y=-\sin\phi\hat x'+\cos\phi \hat y'$.
The effective field of the biaxial model is
\begin{equation}
\begin{aligned}
\mathbf H_\mathrm{eff}=\mathbf H+K_1M_{x'}\hat x'-K_2
M_{z'}\hat z'+\mathrm{Re}(\mathbf he^{-i(\omega t+\phi_2)}),
\end{aligned}
\label{H_eff_bi}
\end{equation}
where $K_1$ and $K_2$ are respectively the dimensionless easy-axis
anisotropy coefficient and hard-axis anisotropy coefficient.
$K_2$ is mainly from the shape anisotropy for a soft magnetic film
so we set $K_2=1$ to acknowledge this fact in this study.

According to Eq. \eqref{llg_eq3}, the linearized LLG Eq. in the
present case becomes
\begin{equation}
\begin{gathered}
-i\omega\mathbf m=-\gamma\mathbf m\times(\mathbf H
+K_1M_{0x'}\hat x'-K_2M_{0z'}\hat z')\\
-\gamma\mathbf M_0\times(\mathbf h+K_1m_{x'}\hat x'
-K_2m_{z'}\hat z')-i\omega\frac{\alpha}{M}\mathbf M_0\times\mathbf m.
\end{gathered}
\label{llg_eq_bi}
\end{equation}
The exact solution of this equation allows us to obtain the expression
of $\tensor\chi$. The non-zero matrix elements of $\tensor\chi$
in the $xyz-$coordinate, according to Eq. \eqref{chi_PQT}, are
$\chi_{xx}=P/(|T|^2-PQ)$, $\chi_{xy}=T/(|T|^2-PQ)$,
$\chi_{yx}=T^\ast/(|T|^2-PQ)$ and $\chi_{xy}=Q/(|T|^2-PQ)$.
$P$, $Q$ and $T$ for our biaxial model are
\begin{equation}
\begin{aligned}
P &= K_1\sin^2\phi-\frac{H_\mathrm{dc}}{M}
+i\alpha\frac{\omega}{\gamma M}\\
Q &= K_1\cos^2\theta\cos^2\phi-K_2\sin^2\theta
-\frac{H_\mathrm{dc}}{M}+i\alpha\frac{\omega}{\gamma M}\\
T &= \frac{K_1}{2}\cos\theta\sin2\phi+i\frac{\omega}{\gamma M},\\
\end{aligned}
\label{chi_bi}
\end{equation}
where the effective dc-field is
$\mathbf H_\mathrm{dc}=\mathbf H+K_1M_{0x'}\hat x'-K_2M_{0z'}\hat z'$.

In our verification of the universal expressions of Eqs. \eqref{chi_f} and
\eqref{chi_H2} below, the model parameters are chosen as
$M=8.0\times10^5$ A/m and $\alpha=0.008$. The value of $K_1$ and field
$\mathbf H$ will vary to model different materials and experimental
configurations. The phase difference $\phi_1-\phi_2$ between the rf
current density $\mathbf j$ (along $\hat y'$) and the rf magnetic field
$\mathbf h$ (along $\hat x'$) is treated as an input parameter, and the
dc-voltage is measured along the $x'-$direction.

In Section II.B, we showed that only a few numbers are needed to fully
characterize $\tensor\chi$, no matter how complicated that a magnetic
anisotropy of a sample might be. In order to verify this assertion,
the comparison between the exact expression of $\tensor\chi$ and its
universal form is given below for two experimental setups in which
static magnetic field $\mathbf H$ is applied either out of magnetic
film in the $y'z'-$plane or in the film plane ($x'y'-$plane).
The resultant dc-voltage lineshapes calculated from the universal form
of $\tensor\chi$ are also compared with numerical simulations based on
the LLG equation and the generalized Ohm's law.

\subsection{Case of $\mathbf H$ in the $y'z'-$plane}
\begin{figure*}
\centering
\includegraphics[width=0.86\textwidth]{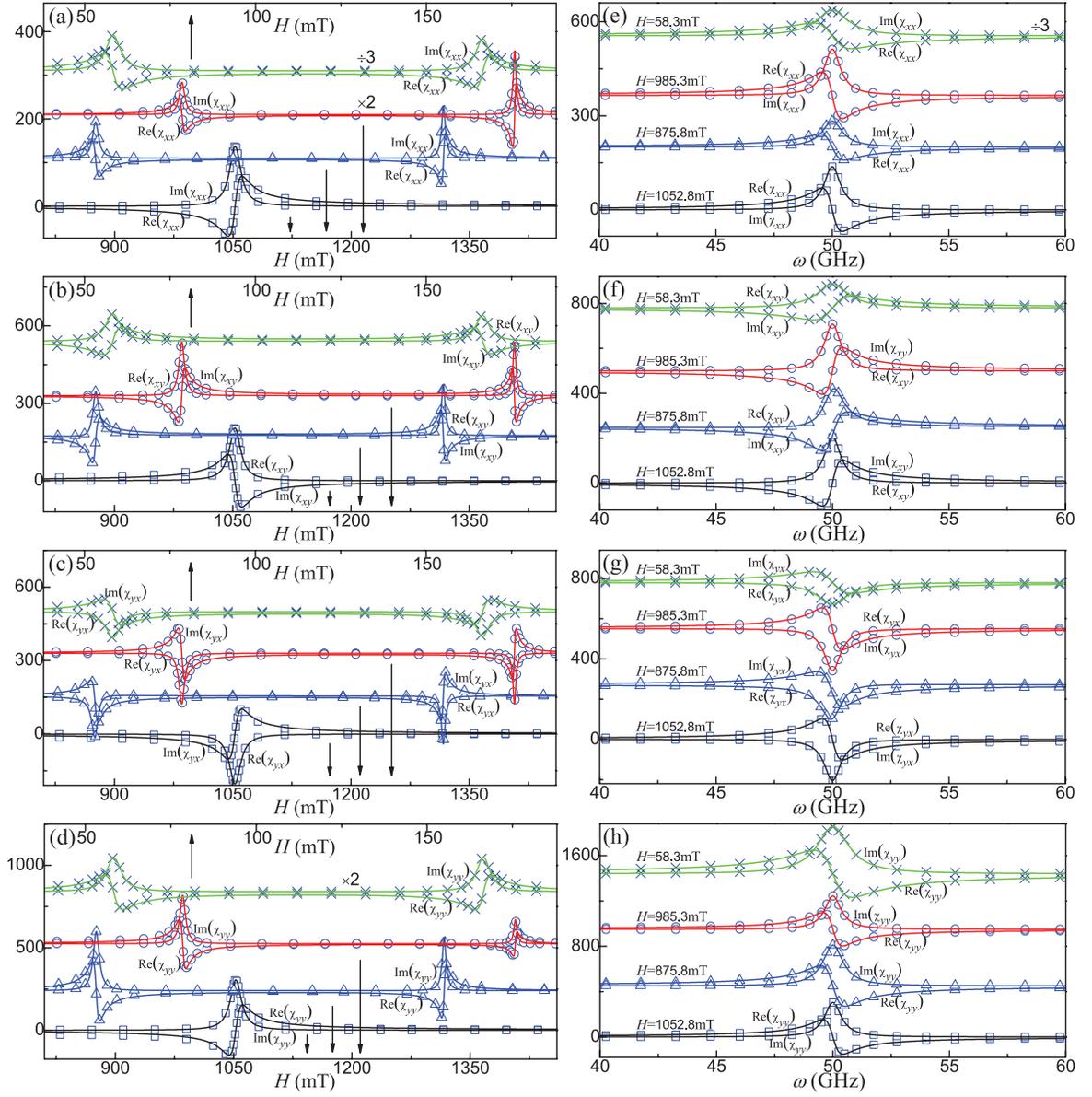}\\
\caption{
The real and imaginary parts of $\chi_{xx}$, $\chi_{xy}$, $\chi_{yx}$
and $\chi_{yy}$ as functions of $H$ at a fixed frequency $\omega=50$
GHz (\textbf a-\textbf d); or as functions of $\omega$ at fixed field
(\textbf e-\textbf h) with $K_1=0$, $K_2=1$, $\theta_H=5.0^o$ and
$\phi_H=90.0^o$ (squares);
or with $K_1=0.2$, $K_2=1$, $\theta_H=5.0^o$ and $\phi_H=90.0^o$ (triangles);
or with $K_1=0.2$, $K_2=1$, and $\mathbf H$ along the $z'$ axis (circles);
or with $K_1=0.1$, $K_2=1$, $\theta_H=90.0^o$ and $\phi_H=87.0^o$ (cross).
Symbols are the exact results from Eq. \eqref{chi_bi} and solid curves
are the fittings to the universal forms of Eqs. \eqref{chi_H2}
(\textbf a-\textbf d) and \eqref{chi_f} (\textbf e-\textbf h).
Curves are vertically offset (from the top down) 310, 210,110, 0 (a);
540, 330, 175, 0 (b); 500, 330, 155, 0 (c); 840, 525, 240, 0 (d);
555, 365, 200, 0 (e); 780, 500, 250 0 (f); 780, 550, 270, 0 (g);
1440, 950, 450, 0 (h) for a better view.
}
\label{chi1}
\end{figure*}

\begin{figure}
\centering
\includegraphics[width=0.45\textwidth]{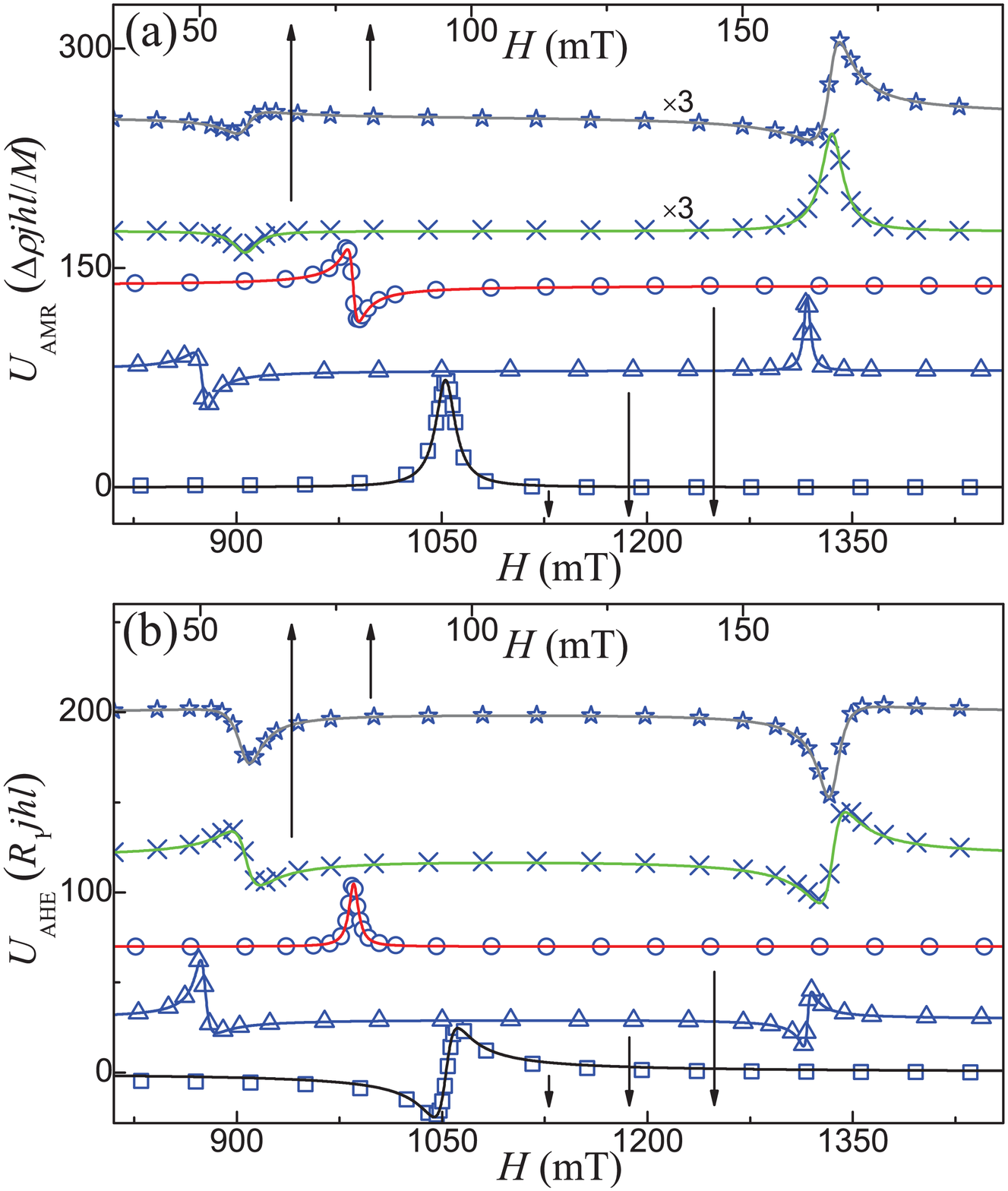}\\
\caption{The field dependence of the dc-voltage due to the AMR contribution
(\textbf a) and the AHE contribution (\textbf b) at $\omega=50$ GHz.
Model parameters are $K_1=0$, $K_2=1$, $\theta_H=5.0^o$, $\phi_H=90.0^o$
and $\phi_1-\phi_2=-90^o$ (squares); or $K_1=0.2$, $K_2=1$, $\theta_H=5.0^o$,
$\phi_H=90.0^o$ and $\phi_1-\phi_2=-90^o$ (triangles); or $K_1=0.2$, $K_2=1$,
$\mathbf H$ along the $z'-$axis and $\phi_1-\phi_2=-90^o$ (circles); or
$K_1=0.1$, $K_2=1$, $\theta_H=90.0^o$, $\phi_H=87.0^o$ and $\phi_1-\phi_2=-90^o$
(cross); or $K_1=0.1$, $K_2=1$, $\theta_H=90.0^o$, $\phi_H=87.0^o$ and
$\phi_1-\phi_2=-30^o$ (stars).
Symbols are the exact results from the LLG equation and the generalized
Ohm's law and solid curves are the universal expression of Eqs. \eqref
{dc_AMR_H2} (\textbf a) and \eqref{dc_AHE_H2} (\textbf b) with $A$'s given by
Eq. \eqref{A} and fitting parameters for the corresponding curves in Fig.
\ref{chi1}.
Curves are vertically offset (from the top down) 254, 175, 138, 80, 0 (a);
200, 120, 70, 30, 0 (b) for a better view.
}
\label{dc}
\end{figure}

In the first example, we set $K_1=0.0$ ($K_2=1$), $\theta_H=5.0^o$,
$\phi_H=90^o$ and $\phi_1-\phi_2=-90^o$. $\mathbf M_0$ is in the $y'z'$-plane
with the polar angle of $\theta=28.7^o$, non-collinear with $\mathbf H$.
Obviously, the system is symmetric about the $y'z'-$plane so that
$\chi_{xy}=-\chi_{yx}$. FMR occurs at $H=1052.8$ mT for $\omega=50$ GHz.
The resonance frequency $\omega_0$ increases with $H$.
The exact results of both real and imaginary parts of
$\chi_{xx}$, $\chi_{xy}$, $\chi_{yx}$ and $\chi_{yy}$ are denoted
by squares in Figs. \ref{chi1}(a)-\ref{chi1}(h).
Figures \ref{chi1}(a)-\ref{chi1}(d) are $\chi$'s as functions of $H$ at a
fixed frequency $\omega=50$ GHz while Figs. \ref{chi1}(e)-\ref{chi1}(h)
as functions of $\omega$ at a fixed field $H=1052.8$ mT.
The black curves are the universal expressions of Eqs. \eqref{chi_H2}
(for Figs. \ref{chi1}(a)-\ref{chi1}(d)) and \eqref{chi_f} (for Figs.
\ref{chi1}(e)-\ref{chi1}(h)) with fitting parameters of $\Gamma=0.87$ GHz,
$\Gamma_1=17.4$ mT, $\beta/|\beta|=1$, $C_1=137.4$, $C_2=0.0$, $C_3=204.6$
and $C_4=304.8$. Curves in this example are either Lorentzian or
$D$ functions according to Eqs. \eqref{chi_H2} and \eqref{chi_f}.
The perfect agreement demonstrates the validity of the universal expressions.
This simple example shows that $\mathbf M_0$ and $\mathbf H$ are not
collinear with each other even in a film made of zero-anisotropy ($K_1=0$)
magnetic materials, in contrast to the popular assumption that
$\mathbf M_0$ is always along $\mathbf H$ for soft magnetic films such
as Permalloy and Yttrium iron garnet.

The dc-voltage is computed from the LLG equation and the generalized Ohm's law.
According to Eqs. \eqref{dc_AMR_chi} and \eqref{dc_AHE_chi},
the dc-voltage due to the AMR involves only Im$(\chi_{yy})$, which
is a Lorentzian function in $H$, while the dc-voltage due to the
AHE involves only Im$(\chi_{xy})$, which is a $D$ function.
According to Eq. \eqref{A}, the above fitting values of $C'$s result in
$A_1=73.3$ and $A_2=0.0$, in units of $\Delta\rho jhl/M$, and $A_3=0.0$
and $A_4=-49.2$, in units of $R_1 jhl$, where $j$ and $h$ are respectively
the amplitudes of the rf current density and the rf magnetic field.
The $H$-dependence of the predicted dc-voltage due to the AMR and AHE are
plotted in Figs. \ref{dc}(a) and \ref{dc}(b) as black curves that describe
perfectly the exact results (squares) from the LLG equation and the
generalized Ohm's law.

\begin{figure}
\centering
\includegraphics[width=0.36\textwidth]{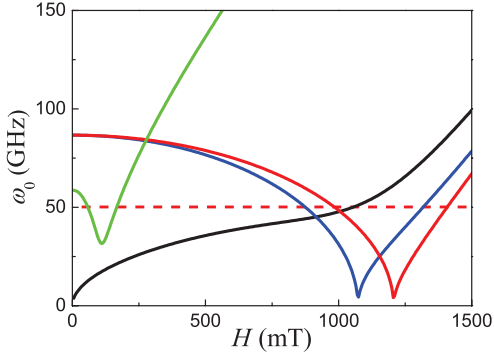}\\
\caption{The $H-$dependence of $\omega_0$ for
$K_1=0.0$, $\theta_H=5.0^o$ and $\phi_H=90.0^o$ (black curves);
or $K_1=0.2$, $\theta_H=5.0^o$ and $\phi_H=90.0^o$ (blue curves);
or $K_1=0.2$, $\mathbf H$ along the $z'-$axis (red curves);
or $K_1=0.1$, $\theta_H=90.0^o$ and $\phi_H=87.0^o$ (green curves).
The dashed line is $\omega=50$ GHz.
}
\label{fH_curve}
\end{figure}

To show the importance of a small non-zero magnetic anisotropy, we
introduce a small $K_1=0.2$ to the first example while keep all other
parameters unchange. Due to the finite easy-axis
anisotropy $K_1$, $\mathbf M_0$ is pushed out of the $y'z'-$plane if
$H$ is not excessive large, clearly non-collinear with $\mathbf H$.
The system is not symmetric about either the $xz-$ or $yz-$plane.
Different from the first example, $\chi_{xy}$ and $\chi_{yx}$ are not
opposite to each other. The exact results of both real and imaginary
parts of $\chi_{xx}$, $\chi_{xy}$, $\chi_{yx}$ and $\chi_{yy}$
are denoted by triangles in Figs. \ref{chi1}(a)-\ref{chi1}(h).
Figures \ref{chi1}(a)-\ref{chi1}(d) are $\chi$'s as functions of $H$
at a fixed frequency $\omega=50$ GHz while Figs. \ref{chi1}(e)-\ref
{chi1}(h) as functions of $\omega$ at a fixed field $H=875.8$ mT.
Two FMR peaks appear in Figs. \ref{chi1}(a)-\ref{chi1}(d) while
there is only one peak in Figs. \ref{chi1}(e)-\ref{chi1}(h).
The direction of $\mathbf M_0$ varies with $H$ unless $\mathbf M_0$ is
collinear with $\mathbf H$ at high values of $H$. As a result, the
effective field also varies with $H$. The effective field is not a
monotonic function of $H$. Thus, $\omega_0(H)$ is also not a monotonic
function of $H$ as shown by the blue curve in Fig. \ref{fH_curve}.
At microwave frequency $\omega=50$ GHz, there are two FMR peaks, or two
values of $H$ that satisfy $\omega_0(H)=\omega$ as shown by the dashed
line of $\omega_0(H)=50$ GHz in Fig. \ref{fH_curve} that cross with blue
curve at $H=875.8$ mT, and $H=1317.0$ mT. For a comparison, the monotonic
$\omega_0-H$ behavior of the first example is also plotted as the black
curve in the figure. At the first peak ($H_0=875.8$ mT), $\mathbf M_0$
lies with $\theta=44.0^o$ and $\phi=33.2^o$,
non-collinear with $\mathbf H$. The second peak occurs at $H_0=1317.0$ mT,
and $\mathbf M_0$ lies in the $y'z'-$plane with $\theta=18.2^o$. The blue
curves in Figs. \ref{chi1}(a)-\ref{chi1}(h) are the universal expressions
of Eqs. \eqref{chi_H2} for (a)-(d) and \eqref{chi_f} for (e)-(h).
For the curves in Figs. \ref{chi1}(a)-\ref{chi1}(d), the fitting parameters
around the first (second) peak are $\Gamma_1=9.2$ mT, $\beta/|\beta|=-1$,
$C_1=83.0$, $C_2=39.5$, $C_3=168.1$ and $C_4=359.2$ ($\Gamma_1=6.0$ mT,
$\beta/|\beta|=1$, $C_1=118.9$, $C_2=0.0$, $C_3=196.1$ and $C_4=323.2$).
For the curves in Figs. \ref{chi1}(e)-\ref{chi1}(h), the fitting
parameters are $\Gamma=1.06$ GHz and the same $C$-values
as those for the first peak in Figs. \ref{chi1}(a)-\ref{chi1}(d).
Different from the first example, the real and imaginary parts of
$\chi_{xy}$ and $\chi_{yx}$ around the first peak are asymmetric functions.
The perfect agreement verifies the validity of the universal expressions.

The numerical results of the dc-voltage due to the AMR and AHE, obtained
from the LLG equation and the generalized Ohm's law, are denoted by
the triangles in Figs. \ref{dc}(a) and \ref{dc}(b), respectively.
Both FMR peaks generate dc-voltage signals.
According to Eq. \eqref{A} and using the fitting parameters found early
for each peak, $A$'s are $A_1=-13.8$, $A_2=35.1$ (in units of
$\Delta\rho jhl/M$), $A_3=-24.8$ and $A_4=-31.9$ (in units of $R_1 jhl$)
for the first FMR peak, and $A_1=49.7$, $A_2=0.0$ (in units of
$\Delta\rho jhl/M$), $A_3=0.0$ and $A_4=-30.1$ (in units of $R_1 jhl$)
for the second FMR peak. The predicted dc-voltage due to the AMR and AHE
(blue curves in Figs. \ref{dc}(a) and \ref{dc}(b)) describe perfectly
the exact numerical results (triangles).
For the first peak, since each of $\mathbf j$, $\mathbf h$ and $\mathbf l$
is out of the $xz-$ or $yz-$plane, the dc-voltage due to both the AMR and
AHE depends on Im$(\chi_{xx})$, Im$(\chi_{xy})$, Im$(\chi_{yx})$ and
Im$(\chi_{yy})$ according to Eqs. \eqref{dc_AMR_chi} and \eqref{dc_AHE_chi}.
Im$(\chi_{xx})$ and Im$(\chi_{yy})$ (Im$(\chi_{xy})$
and Im$(\chi_{yx})$) are asymmetric (symmetric) about FMR peak.
Consequently, the dc-voltage lineshapes due to both the AMR and AHE are
asymmetric. For the second peak, since $\mathbf j$ is in the $xz-$plane
and $\mathbf h$ and $\mathbf l$ are along the $y-$axis, the dc-voltage
due to the AMR relates only to Im$(\chi_{yy})$ according to
Eq. \eqref{dc_AMR_chi} and that due to the AHE relates only to
Im$(\chi_{xy})$ according to Eq. \eqref{dc_AHE_chi}.
Im$(\chi_{yy})$ (Im$(\chi_{xy})$) is symmetric (antisymmetric) about
FMR peak. Consequently, the dc-voltage lineshapes about the second
peak have the same symmetry as those in the first example.

In the third example, we set $K_1=0.2$ ($K_2=1$), $\mathbf H$ along the
$z'-$axis and $\phi_1-\phi_2=-90^o$. $\mathbf M_0$ lies in the $x'z'-$plane
so that the system is symmetric about the $x'z'-$plane (also the $xz-$plane).
Like that in the first example, $\chi_{xy}$ and $\chi_{yx}$ are
opposite to each other. The exact results of both real and imaginary
parts of $\chi_{xx}$, $\chi_{xy}$, $\chi_{yx}$ and $\chi_{yy}$
are denoted by circles in Figs. \ref{chi1}(a)-\ref{chi1}(h).
The $H-$dependences at a fixed frequency of $\omega=50$ GHz are in Figs.
\ref{chi1}(a)-\ref{chi1}(d), and the $\omega-$dependences at a fixed
field of $H=985.3$ mT are in Figs. \ref{chi1}(e)-\ref{chi1}(h).
Two FMR peaks appear in Figs. \ref{chi1}(a)-\ref{chi1}(d) while there
is only one peak in Figs. \ref{chi1}(e)-\ref{chi1}(h).
The two peaks are due to the non-monotonic of the effective magnetic
field in $H$, resulting in multiple solutions (of $H$) which satisfy
$\omega_0(H)=\omega$. The non-monotonic behavior of $\omega_0$ vs. $H$
is shown by the red curve in Fig. \ref{fH_curve}.
In the current case of $\omega=50$ GHz,
there are two resonance fields ($985.3$ mT and $1407.3$ mT).
The first peak occurs at $H_0=985.3$ mT, and $\mathbf M_0$ lies in
the $x'z'-$plane with $\theta=34.9^o$, non-collinear with $\mathbf H$.
The second peak occurs at $H_0=1407.3$ mT, and $\mathbf M_0$ is along
the $z'-$axis, collinear with $\mathbf H$.
The red curves in Figs. \ref{chi1}(a)-\ref{chi1}(h) are the universal
expressions of Eq. \eqref{chi_H2} for (a)-(d) and Eq. \eqref{chi_f}
for (e)-(h). For the curves in Figs.
\ref{chi1}(a)-\ref{chi1}(d), the fitting parameters around the first
(second) peak are $\Gamma_1=8.3$ mT, $\beta/|\beta|=-1$, $C_1=147.5$,
$C_2=0.0$, $C_3=208.5$ and $C_4=294.7$ ($\Gamma_1=4.6$ mT,
$\beta/|\beta|=1$, $C_1=294.8$, $C_2=0.0$, $C_3=208.4$ and $C_4=147.4$).
For the curves in Figs. \ref{chi1}(e)-\ref{chi1}(h), the fitting
parameters are $\Gamma=0.85$ GHz and the same $C$-values
as those for the first peak in Figs. \ref{chi1}(a)-\ref{chi1}(d).
Like the first example, each curve around FMR is either a Lorentzian or
$D$ function. The validity of the universal expressions are
shown by the perfect agreement.

The exact numerical results of the dc-voltage due to the AMR and AHE are
computed from the LLG equation and the generalized Ohm's law, and they are
denoted by the circles in Figs. \ref{dc}(a) and \ref{dc}(b), respectively.
Interestingly, only one peak, corresponding to the first peak in $\chi$'s,
appears. The second FMR peak does not generate dc-voltage signals because
$\mathbf M_0$ is along the $z'-$axis and the $x'y'z'-$coordinate and the
$xyz-$coordinate are coincident with each other. Thus, $l_z=0$, $j_z=0$
and $\mathbf j \times \mathbf m$ is perpendicular to $\mathbf l$, and,
according to Eqs. \eqref{dc_AMR} and \eqref{dc_AHE}, $U_\mathrm{AMR}=U_\mathrm{AHE}=0$.
According to Eq. \eqref{A} and using the fitting parameters found early
for the first peak, $A$'s are $A_1=0.0$ and $A_2=-49.1$, in units
of $\Delta\rho jhl/M$, and $A_3=34.8$ and $A_4=0.0$, in units of $R_1 jhl$.
The predicted dc-voltage due to the AMR and AHE (red curves in Figs.
\ref{dc}(a) and \ref{dc}(b)) agrees again perfectly with the exact
results (circles) around the first FMR peak.
Since $\mathbf j$ is along the $y$-axis and $\mathbf h$ and $\mathbf l$ are
in the $xz-$plane in this case, the dc-voltage due to the AMR depends on
Im$(\chi_{yx})$ according to Eq. \eqref{dc_AMR_chi} and that due to the AHE
depends only on Im$(\chi_{xx})$ according to Eq. \eqref{dc_AHE_chi}.
Im$(\chi_{yx})$ (Im$(\chi_{xx})$) is antisymmetric (symmetric) about FMR peak.
Consequently, the dc-voltage lineshapes have the opposite symmetry to the
first example. In summary, these examples show that the dc-voltage lineshapes
can change from symmetric to antisymmetric or even vanish due to magnetic
anisotropy.

\subsection{Case of $\mathbf H$ in the $x'y'-$plane}

In the forth example, we set $K_1=0.1$ ($K_2=1$), $\theta_H=90.0^o$,
$\phi_H=87.0^o$ and $\phi_1-\phi_2=-90^o$. $\mathbf M_0$ lies in the
$x'y'-$plane in this case so that the system is symmetric about the
$x'y'-$plane (also the $yz-$plane). With this symmetry, $\chi_{xy}$
and $\chi_{yx}$ are opposite to each other.
The exact results of both real and imaginary parts of $\chi_{xx}$,
$\chi_{xy}$, $\chi_{yx}$ and $\chi_{yy}$ are denoted by cross in Figs.
\ref{chi1}(a)-\ref{chi1}(h) in which Fig.  \ref{chi1}(a)-\ref{chi1}(d)
(\ref{chi1}(e)-\ref{chi1}(h)) are $H-$dependence  ($\omega-$dependence)
at a fixed frequency of $\omega=50$ GHz (field of $H=58.3$ mT).
Two FMR peaks appear in Figs. \ref{chi1}(a)-\ref{chi1}(d) while there
is only one peak in Figs. \ref{chi1}(e)-\ref{chi1}(h).
For the same reason as that for the second example, the two peaks are due
to the non-monotonic of the effective magnetic field in $H$, resulting
in multiple solutions (of $H$) which satisfy $\omega_0(H)=\omega$.
The non-monotonic behavior of $\omega_0$ vs. $H$ is shown by the green
curve in Fig. \ref{fH_curve}.
The first peak occurs at $H_0=58.3$ mT, and the corresponding $\mathbf M_0$
lies in the $x'y'-$plane with $\phi=33.9^o$, non-collinear with $\mathbf H$.
The second peak occurs at $H_0=166.5$ mT at which $\mathbf M_0$ lies in
the $x'y'-$plane with $\phi=82.5^o$, also non-collinear with $\mathbf H$.
The green curves in Figs. \ref{chi1}(a)-\ref{chi1}(h) are the universal
expressions of Eq. \eqref{chi_H2} for (a)-(d) and Eq. \eqref{chi_f} for
(e)-(h). For the curves in Figs. \ref{chi1}(a)-\ref{chi1}(d), the fitting
parameters around the first (second) peak are $\Gamma_1=5.2$ mT, $\beta/|
\beta|=-1$, $C_1=27.2$, $C_2=0.0$, $C_3=106.3$ and $C_4=415.0$ ($\Gamma_1=4.8$
mT, $\beta/|\beta|=1$, $C_1=24.5$, $C_2=0.0$, $C_3=101.2$ and $C_4=417.7$).
For the curves in Figs. \ref{chi1}(e)-\ref{chi1}(h), the fitting parameters
are $\Gamma=1.66$ GHz and the same $C$-values as those for the first peak
in Figs. \ref{chi1}(a)-\ref{chi1}(d). Similr to that of the first example,
each curve around FMR is either a Lorentzian or $D$ function.
The perfect agreement verifies the validity of the universal expressions.

The exact numerical results of the dc-voltage due to the AMR and AHE,
obtained from the LLG equation and the generalized Ohm's law, are
denoted by the cross in Figs. \ref{dc}(a) and \ref{dc}(b), respectively.
Both FMR peaks generate dc-voltage signals in this case.
According to Eq. \eqref{A} and using the fitting parameters found early
for the first (second) peak, $A$'s are $A_1=-43.5$ and $A_2=0.0$, in units
of $\Delta\rho jhl/M$, and $A_3=0.0$ and $A_4=-29.7$, in units of $R_1 jhl$
($A_1=200.1$ and $A_2=0.0$, in units of $\Delta\rho jhl/M$,
and $A_3=0.0$ and $A_4=-50.2$, in units of $R_1 jhl$).
The predicted dc-voltage due to the AMR and AHE (green curves in Figs.
\ref{dc}(a) and \ref{dc}(b)) describes perfectly the exact results (cross).
Since $\mathbf j$, $\mathbf h$ and $\mathbf l$ are all in the $yz-$plane,
the dc-voltage around each FMR peak due to the AMR relates only to
Im$(\chi_{yy})$ according to Eq. \eqref{dc_AMR_chi} and that due to the AHE
relates only to Im$(\chi_{xy})$ according to Eq. \eqref{dc_AHE_chi}.
Im$(\chi_{yy})$ (Im$(\chi_{xy})$) is symmetric (antisymmetric) about FMR peak.
Therefore, the dc-voltage lineshapes in this configuration have the same
symmetries of Im$(\chi_{yy})$ and Im$(\chi_{xy})$.

To show the importance of the phase difference $\phi_1-\phi_2$ between
rf current density and rf magnetic field, we change this phase difference
in the forth example to $\phi_1-\phi_2=-30^o$ while keep all other
parameters in the forth example unchange. Obviously, the dynamic magnetic
susceptibility matrix $\tensor\chi$ will not change. However, the symmetry
of dc-voltage lineshapes change due to a different $\phi_1-\phi_2$.
The exact numerical results of the dc-voltage due to the AMR and AHE are
denoted by the stars in Figs. \ref{dc}(a) and \ref{dc}(b), respectively.
Similar to the previous example, both FMR peaks generate dc-voltage signals
in this case.
According to Eq. \eqref{A} and using the fitting parameters found early
for the first (second) peak, $A$'s are $A_1=-21.7$ and $A_2=37.7$, in units
of $\Delta\rho jhl/M$, and $A_3=-25.7$ and $A_4=-14.9$, in units of $R_1 jhl$
($A_1=100.1$ and $A_2=-173.3$, in units of $\Delta\rho jhl/M$,
and $A_3=-43.4$ and $A_4=-25.1$, in units of $R_1 jhl$).
The predicted dc-voltage due to the AMR and AHE (gray curves in Figs.
\ref{dc}(a) and \ref{dc}(b)) describes perfectly the exact results (stars).
Since $\mathbf j$, $\mathbf h$ and $\mathbf l$ are all in the $yz-$plane
and $\phi_1-\phi_2=-30^o$,
the dc-voltage around each FMR peak due to the AMR relates to both
Re$(\chi_{yy})$ and Im$(\chi_{yy})$ according to Eq. \eqref{dc_AMR_chi}
and that due to the AHE relates to both Re$(\chi_{xy})$ and Im$(\chi_{xy})$
according to Eq. \eqref{dc_AHE_chi}.
Re$(\chi_{xy})$ and Im$(\chi_{yy})$ (Re$(\chi_{yy})$ and Im$(\chi_{xy})$)
are symmetric (antisymmetric) about FMR peak.
Therefore, the dc-voltage lineshapes in this configuration are asymmetric.

\subsection{Non-resonance peaks}

\begin{figure}
\centering
\includegraphics[width=0.5\textwidth]{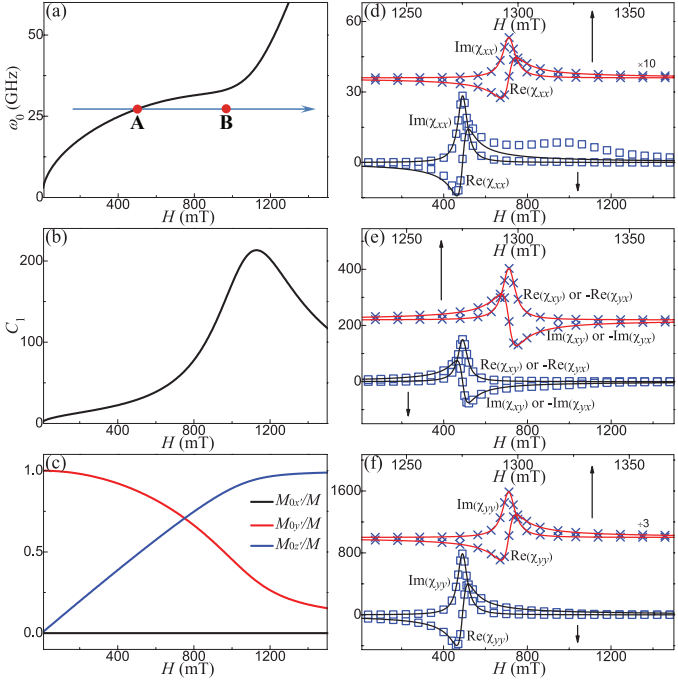}\\
\caption{The model parameters are $K_1=0.0$ ($K_2=1$), $\theta_H=3.0^o$
and $\phi_H=90.0^o$. (\textbf a) $H-$dependence of $\omega_0$.
$\omega_0$ does not change much in the range of 400 mT $<H<$ 900 mT.
At $\omega=27$ GHz (blue arrow line), $\beta = 0.02 $ GHz/mT.
Point A and Point B are respectively the resonance and non-resonance
peaks of Re$(\chi_{xx}(H))$. (\textbf b) $H-$dependence of $C_1(H)$.
(\textbf c) $H-$dependence of $M_{0x'}$, $M_{0y'}$ and $M_{0z'}$.
(\textbf d)-(\textbf f) The real and imaginary parts of $\chi_{xx}$,
$\chi_{xy}$ (or $-\chi_{yx}$) and $\chi_{yy}$ as functions of $H$.
Exact results are denoted by symbols, squares for $\omega=27$ GHz and
cross for $\omega=60$ GHz. Solid curves are universal expression of
Eq. \eqref{chi_H2} with the best fitting parameters, black for
$\omega=27$ GHz and red for $\omega=60$ GHz.
Top curves are vertically offset 36 (d), 220 (e)
and 1000 (f) for a better view.
}
\label{broad}
\end{figure}

So far, we have demonstrated the validity of universal expressions of
Eqs. \eqref{chi_H2} and \eqref{chi_f} for $\tensor\chi$, and Eqs.
\eqref{dc_AMR_chi} and \eqref{dc_AHE_chi} for dc-voltage near FMR in a
number of examples as long as the approximation of Eq. \eqref{approx}
is good. When Eq. \eqref{approx} is not good, some of the matrix
elements of $\tensor\chi$ do not follow Eq. \eqref{chi_H2}, and an
extra non-resonance broad peak can even appear in some of the matrix
elements near a true FMR peak in such a circumstance.
To see a clear example, we set $K_1=0.0$ ($K_2=1$), $\theta_H=3.0^o$,
$\phi_H=90.0^o$ and $\omega=27$ GHz. $\mathbf M_0$ lies in the $y'z'-$plane,
and the system is symmetric about the $y'z'-$plane (also the $xz-$plane)
so that $\chi_{xy}$ and $\chi_{yx}$ are opposite to each other.
Figure 6(a) shows how FMR resonance frequency $\omega_0$ changes with $H$.
Around relative low frequencies ($\simeq 27$ GHz), $\omega_0$ is
not very sensitive to $H$ with a very small $\beta=0.02$ GHz/mT.
For both Lorentzian function $L$ and the corresponding $D$ function
which decay as a power-law, the resonance region should be several
peak widths whose value is $\Gamma \simeq 1$ GHz in current example.
This means that the field in the resonance region can vary from about
$300$ mT ($\omega_0\sim 23$ GHz) to $1000$ mT ($\omega_0\sim 31$ GHz).
In this field range, $C_1$ increases by more than ten times
as shown in Fig. 6(b), and the direction of $\mathbf M_0$, around the
$y'-$axis at 300 mT, moves to near the $z'-$axis at 1000 mT (Fig. 6(c)).
Thus, one should not expect our $\tensor\chi$ expression of Eq.
\eqref{chi_H2} good any more.

The exact results of both real and imaginary parts of $\chi_{xx}$,
$\chi_{xy} (=-\chi_{yx})$ and $\chi_{yy}$ as functions of $H$ at
fixed frequency $\omega=27$ GHz (corresponding to point A in Fig. 6(a))
are plotted as squares in Figs. \ref{broad}(d)-\ref{broad}(f).
The FMR peak occurs at $H_0=490.4$ mT, and $\mathbf M_0$ lies in the
$y'z'-$plane with $\theta=61.8^o$, non-collinear with $\mathbf H$.
The black curves in Figs. \ref{broad}(d)-\ref{broad}(f) are the
universal expression of Eq. \eqref{chi_H2} with the fitting parameters
of $\Gamma_1=54.9$ mT, $\beta/|\beta|=1$, $C_1=28.4$, $C_2=0.0$,
$C_3=149.8$ and $C_4=790.5$. The deviation between the exact results
and best fit to Eq. \eqref{chi_H2} at higher $H$ ($>H_0$) is obvious,
especially for Re$(\chi_{xx})$ whose exact result has an extra
non-resonance broad peak at 970 mT.


For a comparison, we also plot the $\chi$'s for $\omega=60$ GHz at which
$\beta$ is larger and the approximation of Eq. \eqref{approx} is good.
The exact results of both real and imaginary parts of $\chi_{xx}$,
$\chi_{xy}$ (or $-\chi_{yx}$) and $\chi_{yy}$ as functions of
$H$ are denoted by cross in Figs. \ref{broad}(d)-\ref{broad}(f).
The FMR peak occurs at
$H_0=1295.9$ mT, and $\mathbf M_0$ lies in the $y'z'-$plane with
$\theta=12.3^o$, non-collinear with $\mathbf H$.
The red curves in Figs. \ref{broad}(d)-\ref{broad}(f) are the universal
expression of Eq. \eqref{chi_H2} with the fitting parameters of
$\Gamma_1=6.8$ mT, $\beta/|\beta|=1$, $C_1=172.0$, $C_2=0.0$,
$C_3=183.8$ and $C_4=196.5$. Different from the low frequency case,
Eq. \eqref{chi_H2} describes well the exact results again.

\section{Conclusions}

In this work, we show that both the dynamic magnetic susceptibility and
the dc-voltage lineshapes in the electrical detection of FMR takes
universal expressions near FMR for an arbitrary magnetic anisotropy.
The magnetic anisotropy can affect values of those parameters that
uniquely define the universal functions, but it cannot change the forms.
Our explicit examples show that both the real and imaginary parts of
$\chi_{xy}$ and $\chi_{yx}$ can be asymmetric, symmetric and antisymmetric,
depending on the magnetic anisotropy and experimental setup.
It is already known that interfacial interaction can change the magnetic
anisotropy \cite{sakurai, tsujikawa, zli} in multilayer structures.
Thus the dc-voltage signal, including its shape and symmetry, can be
very different for a multilayer film and for a single magnetic layer.

In conclusion, the universal form of the dynamic magnetic susceptibility
matrix is found by using the causality principle and the assumption that
the frequency dependence of a usual microwave absorption lineshape around
a FMR is Lorentzian. The dynamic magnetic susceptibility is not,
in general, the Polder tensor that is normally assumed in the literature.
The non-Polder tensor of the dynamic magnetic susceptibility has important
effects on the dc-voltage lineshape in the electrical detection of a FMR.
It is also found that the linear response coefficient of the
magnetization to the total local rf field (sum of the applied external rf
field and the internal rf field due to the precessing magnetization, a
quantity cannot be measured directly) is a Polder tensor. This finding
may explain why the Polder tensor was so widely misused in analyses.
Although the dynamic magnetic susceptibility and dc-voltage near the FMR
should depend on, in principle, the magnetic anisotropy, we show that
they are fully determined by a few parameters instead of a function.
These parameters can be experimentally determined by traditional
microwave absorption experiments. Our results provide a reliable
way to extract the dc-voltage induced by spin pumping and the ISHE.
Furthermore, the non-monotonic behavior of the total effective magnetic
field in external field may lead to multiple FMR peaks in field at a
constant microwave frequency while there is only a single frequency
peak at a fixed static field. We also point out that insensitivity of
resonance frequency to the magnetic field may lead to another broad peak
which does not correspond to FMR. It raises an alarm on proper
interpretation of detected FMR field peaks, especially those with broad
peak and unfitable to our universal expressions.

\section{Acknowledgements}
This work was supported by Hong Kong RGC grants (163011151 and 605413),
grant from National Natural Science Fundation of China (11374249) and UESTC.
SSK acknowledge the support of 111 Project grant (B13029).


\end{document}